\documentclass[final,,twoside]{IEEEtran}
\usepackage{fancyhdr}
\usepackage{amsmath,amssymb}
\usepackage{amsfonts}
\usepackage{amsthm}
\usepackage[dvips]{graphicx}
\usepackage{dsfont}
\usepackage{balance}
\usepackage{algpseudocode}
\usepackage{algorithm}

\usepackage{comment}
\usepackage{amsthm}
\usepackage{multicol,multirow}
\usepackage{graphicx,subcaption}
\usepackage{amssymb, amsmath}
\usepackage{xcolor}
\usepackage{cite}
\usepackage{hyperref}
\usepackage{enumerate}
\usepackage{esint}
\usepackage{balance}
\usepackage{textcomp}
\usepackage{bbold}
\usepackage{longtable}
\usepackage{algorithm}
\usepackage{algpseudocode}
\usepackage{cuted}

\graphicspath{{figures/}}
\usepackage{nomencl}
\usepackage{acronym} 
\makenomenclature

\begin{document}

\title{High-Fidelity Coherent-One-Way QKD Simulation Framework for 6G Networks: Bridging Theory and Reality}

% !!!!!!!!!!!TBD!!!!!!!!!!!
\author{Aitor Brazaola-Vicario, Vasileios Kouvakis, 
        Stylianos E. Trevlakis~\IEEEmembership{Member,~IEEE,}
        Alejandra Ruiz, Alexandros-Apostolos A. Boulogeorgos,~\IEEEmembership{Senior Members,~IEEE,} and
        Theodoros Tsiftsis, \IEEEmembership{Senior Members,~IEEE}, Dusit Niyato, \IEEEmembership{Fellow, IEEE}
\thanks{A. Brazaola-Vicario and A. Ruiz is with TECNALIA, Basque Research and Technology Alliance (BRTA), Spain, E-mails: }
\thanks{V. Kouvakis, S. Trevlakis, and T. A. Tsiftis is with the Department of Research and Development, InnoCube P.C., 79, 17is Noemvriou Str., 55534 Thessaloniki, Greece, E-mails: \{kouvakis,trevlakis, tsiftsis\}@innocue.org}
\thanks{A.-A. A. Boulogeorgos is with the Department of Electrical and Computer Engineering, University of Western Macedonia, ZEP Area, 50100 Kozani, Greece, E-mail: aboulogeorgos@uowm.gr}
\thanks{T. A. Tsiftis is also with the Department of Informatics \& Telecommunications, University of Thessaly, Lamia 35100, Greece, E-mail: tsiftsis@uth.gr}
\thanks{D. Niyato is with the College of Computing and Data Science, Nanyang Technological University, Singapore 639798 (e-mail: dniyato@ntu.edu.sg).}
%\corresp{CORRESPONDING AUTHOR: Aitor Brazaola-Vicario (e-mail: aitor.brazaola@tecnalia.com).}
\thanks{This project has received funding from the European Union’s Horizon Europe Framework Programme under grant agreement No 101096456 (NANCY).}}

\markboth{High-Fidelity Coherent-One-Way QKD Simulator: Bridging Theory and Reality}{Brazaola-Vicario \textit{et al.}}

\maketitle
\begin{abstract}
Quantum key distribution (QKD) has been emerged as a promising solution for guaranteeing information-theoretic security.  Inspired by this, a great amount of research effort has been recently put on designing and testing QKD systems as well as articulating preliminary application scenarios. However, due to the considerable high-cost of QKD equipment, a lack of QKD communication system design tools, wide deployment of such systems and networks is challenging. Motivated by this, this paper introduces a QKD communication system design tool. First we articulate key operation elements of the QKD, and explain the feasibility and applicability of coherent-one-way (COW) QKD solutions. Next, we focus on documenting the corresponding simulation framework as well as defining the key performance metrics, i.e., quantum bit error rate (QBER), and secrecy key rate. To verify the accuracy of the simulation framework, we design and deploy a real-world QKD setup. We perform extensive experiments for three deployments of diverse transmission distance in the presence or absence of a QKD eavesdropper. The results reveal an acceptable match between simulations and experiments rendering the simulation framework a suitable tool for QKD communication system design.            
\end{abstract}


\section{Introduction}
\label{sec:introduction}
\IEEEPARstart{A}{s} we move towards next generation communications and networking systems era, security becomes a fundamental system requirement for most verticals~\cite{Je2021,atutxa24,Ntanos2020}. In general two types of security exist. The first type lies on creating cryptographic approaches that require a remarkable amount of time in order to be broken by nowadays processing systems, and is called computational security. The second type is information-theoretic security, which guarantees hard security, or, in other words, unbreakable encryption against adversaries with unlimited computing resources and time~\cite{Rana2022}. 

Despite the paramount usage of computational-security protocol, recent advancements in quantum computing created concerns about the effectiveness and resilience of the aforementioned approaches~\cite{arute2019,debnath2016,gong2021}. On the other hand, the same quantum computing advancements opened the door to new information-theoretic security approaches. Maybe the most indicative example is quantum key distribution (QKD). QKD is a secure communication method that uses quantum mechanics to exchange encryption keys between two entities. It relies on the properties of quantum particles, such as photons, which allow the detection of eavesdropping attempts since any measurement disturbs the system. As a result, QKD provides a theoretically unbreakable method of key sharing~\cite{Cao2022}. 

Inspired by this, a great amount of research effort was put in designing and studying the feasibility of QKD protocols. Two of the most notable protocols are BB84~\cite{Bennet1984} and E91~\cite{Ekert1991}. In BB84, the sender, called Alice, transmits photons polarized in one of four possible states, and the legitimate receiver, i.e. Bob, estimates the polarization using randomly selected bases. By comparing a subset of their measurements, Alice and Bob can detect eavesdropping and establish a shared secret key. On the other hand, E91 protocol is based on the fundamental principles of quantum entanglement and Bell’s theorem. Another interesting protocol, namely G20, was presented by Grosshans and Grangier in 2002~\cite{Grosshans2002}, which implements  Gaussian-modulated  continuous-variable (CV)-QKD relying on coherent states. Coherent-one way (COW) QKD is another protocol of high-interest due to its fair trade-off between low-complexity implementation and high performance in terms of quantum bit error rate (QBER) and secrecy key rate (KeyRate)~\cite{Stucki2005}. 

From the feasibility study point of view, a relatively small number of published contributions provides analytical approaches to quantify the performance of the aforementioned protocols~\cite{Lucamarini2015,Alshaer2021,10214294}. The authors of~\cite{Lucamarini2015} studied security bounds of BB85 QKD fiber systems. An error analysis of E91 QKD assuming free-space optical channel was conducted in~\cite{Alshaer2021}. In~\cite{10214294}, the authors presented an analytical framework for the error and key-rate evaluation of COW-QDK that accounts for the particularities of gigabit passive optical networks for fiber to the home applications. Although the aforementioned contributions have set the stage of analyzing the performance and understanding the parameters that play a key role on the operation of QKD protocols, they have not backed up through experiments. 

On the other hand, recently several in-lab and real-world experimental setups of wired QKD implementations were reported in the literature~\cite{Bahrami2020,Biswas2021,Burdiak2023,Chen2009}. Specifically, in~\cite{Bahrami2020}, the authors studied the performance of QKD applying wavelength division multiplexing over pre-deployed optical fibers. Although the results were promising, there are technical limitations in terms of spontaneous Raman scattering, four-wave mixing, and amplified spontaneous emission, which require rethinking and optimizing the QKD protocol as well as designing predistortion or mitigation approaches. In~\cite{Biswas2021}, the authors reported a BB84 protocol experimental implementation that employs four laser diodes and discussed the impact of laser parameters mismatches. The results of~\cite{Biswas2021} revealed the detrimental impact of laser parameters mismatches, which in turn is translated in the need for developing analog and digital mismatches cancellation or mitigation schemes. In~\cite{Burdiak2023}, a CV-QKD link was demonstrated for backhaul connection.  Finally, the authors of~\cite{Chen2009} documented a wavelength-routing star type QKD network, which can span a metropolis using a commercial backbone optical fiber network without trusted relays. The feasibility of the star-network that employs QKD that was presented in~\cite{Chen2009} was verified through experiments; however, its limits in terms of the number of nodes that can support was not investigated due to the limited available quantum transceivers.  

The development of an appropriate simulation framework is required to bridge the theoretical and development worlds, to design new protocols, to optimize existing ones, to open the door in the developing of pre-distortion and/or mitigation schemes, to allow for QKD network planning without the need of investing in expensive QKD equipment, and to bring to life the concept of software define quantum networking~\cite{Tessinari2023}. Scanning the open literature, someone can identify several open-source simulators that focus on the physical study of the QKD network without accommodating either the particularities of the communication mechanisms or the key management system (KMS) characteristics~(e.g.,~\cite{Aji2021} and references therein). As a consequence, on the one hand, they sacrifice accuracy providing opportunistic results. On the other does not provide the key constraints that need to be followed when adding a new mechanism or optimizing the QKD system. To overcome these limitations and to fill the aforementioned gaps, in this paper, we articulate an experimental-verified simulation framework that is in line with the European Telecommunications Standards Institute (ETSI) interfaces and the corresponding mechanisms~\cite{Dervisevic2024}. In more detail, the contributions of this work are as follows:
\begin{itemize}
    \item We first explain the importance of COW protocol that renders it the most suitable candidate for real-world deployment in next generation networks. Then, we summarize the COW protocol fundamental characteristics and also articulate its operation.
    \item Building upon the fundamental characteristics of the COW protocol as well as the ETSI QKD-related mechanisms, we provide a COW-QKD simulation framework.
    \item We present a real-world experimental setup that is located in TECNALIA premises in Derio (Spain), and we give experimental scenarios that we conducted in order to verify the simulation framework. 
    \item Finally, to demonstrate applicability of the simulation framework, we present simulation results that provides engineering insights and guidelines for optimizing the COW-QKD operation. 
\end{itemize}

The outline of the remainder of this paper is as follows: Section~\ref{sec:background-knowledge} revisits the COW protocol and provides its fundamental characteristics. Section~\ref{sec:simulation-framework} introduces the simulation framework. In Section~\ref{sec:experimental-setup}, the experimental demonstrator, the physical testbed, and the experimental scenarios are presented. The verification of the simulation framework through experimental measurements is provided in Section~\ref{sec:results-and-discussion}. Final remarks and the main message of this contribution is provided in Section~\ref{sec:conclusion}.

%\subsection*{NOMENCLATURE} 
%\input{acro}

%\textcolor{red}{\textbf{DOUBLE CHECK THE NOMENCLATURE}}

\section{Background Knowledge}\label{sec:background-knowledge}
Utilizing decoy states to boost security, H. Zbinden \textit{et al.} developed the COW protocol in 2004 \cite{Stucki2005}, \cite{Ahmed2012}. Since the COW protocol is mostly based on passive optical components, it stands out for simplicity of implementation. Moreover, it is polarization insensitive, which makes it perfect for fiber-based communications free of polarization control devices \cite{Stucki2005}. This architectural benefit allows the COW protocol to link easily into contemporary fiber-optic system. The COW protocol's resistance against photon number splitting (PNS) attacks—a common weakness in many QKD systems—is among its most important benefits. Though its resistance to other kinds of attacks is still under investigation, new experimental implementations have produced positive findings~\cite{Ahmed2012}. For instance, although other studies have reported rates of up to $15$ bps over $250$ km, safe key distribution rates of $2.5$ b/s have been recorded throughout lengths of $150$ km~\cite{Stucki_2009}. Emphasizing simplicity and efficiency, the architecture of the COW protocol offers several advantages. It is an ideal choice for long-distance fiber-based QKD systems that do not require complex polarization control techniques due to its polarization insensitivity and dependence on passive optical components. Furthermore, the protocol's resilience to PNS attacks adds an additional layer of security. The COW protocol's high key generation rates, low implementation complexity, and robust security features make it a strong candidate for real-world quantum communication systems. 

\begin{figure}
    \centering
    \includegraphics[width=1\linewidth]{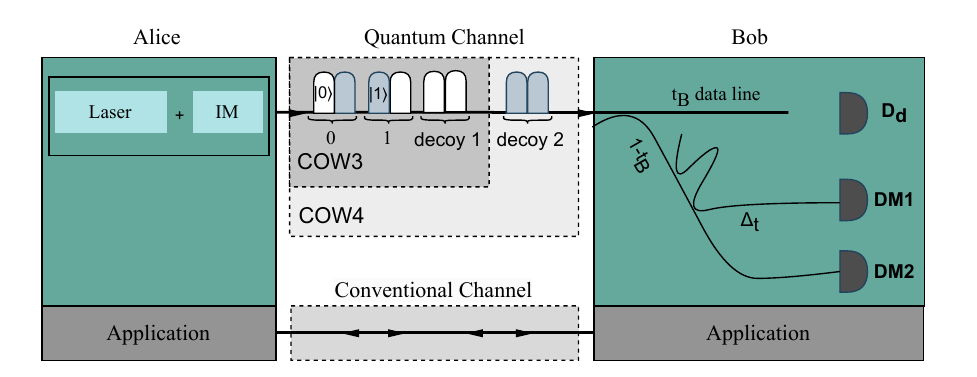}
    \caption{High-level QKD framework.}
    \label{Fig:qkd_arch}
\end{figure}
The fundamental framework of QKD communication process incorporates both quantum and conventional channels that operate in a collaborative fashion. As illustrated in Fig.~\ref{Fig:qkd_arch}, the quantum channel is responsible for the sensitive work of QKD, while the conventional channel supports the essential post-processing session, including key agreement and eavesdropper detection between Alice and Bob. In order to ensure that any unauthorized access is quickly identified and eradicated, this dual channel structure is an essential component of QKD systems. The utilization of the unique physical properties of quantum information carriers enables QKD to be attained, while simultaneously protecting against surveillance. Notice that each attempt by an unauthorized third party to obtain knowledge of the shared key results in a substantial increase in the QBER of the transmitted data. The secure communication rate, when combined with the QBER, is a critical metric for assessing the security and efficacy of a variety of QKD systems~\cite{Buttler1998}. The COW protocol is intended to generate high key rates by utilizing the timing of photon arrivals at the receiver.  According to~\cite{Stucki2005}, the COW-QKD operation is summarized as follows: 

\begin{enumerate}
\item Alice compresses binary bits into time slots before transmitting them to Bob. In a basic configuration without decoy states, the likelihood of communicating a bit of 1 is the same as the probability of broadcasting a bit of 0 (both $50\%$). Alice encodes this logic values by using two-pulse states 
\begin{align}
\left|0_n\right\rangle=|0\rangle_{2 n-1}|\alpha\rangle_{2 n}
\end{align}
and 
\begin{align}
\left|1_n\right\rangle=|\alpha\rangle_{2 n-1}|0\rangle_{2 n}
\end{align}
at two consecutive time windows, i.e., $2n-1$ and $2n$, where $n=1,2, \ldots, N$; with $|0\rangle$ representing the vacuum state and $|\alpha\rangle$ represents the coherent state.  However, for COW4, decoy states with two $|0\rangle$ vacuum states are introduced and the odds change. If the probability of creating a decoy bit is $f$, the remaining probability $(1-f)$ is equally divided between 1 and 0, with each having a probability of 
\begin{align}
p = \frac{1-f}{2(1-f)}.
\end{align}

\item Bob, as the legitimate receiver, uses a beam slitter device with transmittance $t_B$ to passively route incoming states to either the data line or the monitoring line. On the data line (see Fig.~\ref{Fig:qkd_arch}), Bob measures the click time of each signal to determine the logic bit that Alice encoded for round $n$, obtaining the raw key. On the monitoring line, Bob also records which detector clicks in each round. Only the clicks resulting from interference between two pulses from the same round are recorded; while any additional detector's clicks in the round are neglected. In the event that multiple detectors register clicks in a single round, Bob randomly selects one click to record.

\item We assume a passive eavesdropper (Eve) that is able to collect a portion of the light transmitted by Alice, but not perform active attacks. Bob announces in which $n$ round that he recorded a click from the transmitted data. In turn, Alice keeps only the corresponding bits, discarding the rest, in order to form the final raw key. Additionally, Bob also announces his click records from the monitoring line. Alice then calculates the click counts: $t_{0 \alpha}^{M_i}, t_{\alpha 0}^{M_i}, t_{\alpha \alpha}^{M_i}$, and $t_{00}^{M_i}$ (where $i=0$ or 1). Here, $t_u^{M_i}$, for $u=0 \alpha, \alpha 0, \alpha \alpha, 00$, represents the click counts for the states $|0\rangle_{2 n-1}|\alpha\rangle_{2 n}, |\alpha\rangle_{2 n-1}|0\rangle_{2 n}, |\alpha\rangle_{2 n-1}|\alpha\rangle_{2 n}$, and $|0\rangle_{2 n-1}|0\rangle_{2 n}$, respectively, with the superscript $M_i$ indicating clicks on the monitoring line. Using Kato's inequality~\cite{kato1972schrodinger}, Alice estimates the upper bound of the phase error rate $\overline{E_p}$. The QBER $E_z$ is calculated by revealing some bits from the raw key. If either $\overline{E_p}$ or $E_z$ exceeds predefined thresholds, the protocol is aborted.

\item For the final key, error corrections and privacy amplification is applied to increase the possibilities of successful message delivery without the Eve's presence. 
\end{enumerate}

The research presented in this paper was motivated by the need to address the limitations observed in previous studies, which were primarily focused on simulating discrete variable QKD protocols, such as the BB84, or to provide a tool for testing commercial QKD equipment without requiring significant hardware investments. By utilizing existing dark fibres, adhering to ETSI standards, and developing a use case centred on testing applications, this work enables the investigation of the performance of modern QKD protocols, which were inaccessible or impractical in earlier works.

The ``how'' lies in the innovative approach of capturing information using real equipment configured for two variants of the same protocol (COW3 and COW4) and monitoring the performance of the key management system (KMS) used to deliver the keys. This allows for real-time measurements with greater precision than simulations of the protocol alone could achieve.

The ``why'' is rooted in the potential of this experiment to extend the adoption of QKD in conventional use cases that could benefit from the enhanced security provided by this technique, particularly in an era where classical asymmetric cryptography is increasingly threatened by quantum algorithms. Unlike prior research, which was constrained by limited access to commercial hardware and often based on theoretical frameworks, this work offers new opportunities to integrate QKD into more conventional and practical use cases.

\section{Simulation Framework \& Key Performance Metrics}\label{sec:simulation-framework}

This section presents the COW-QKD simulation framework that was designed and developed to assess the performance of QKD systems before their actual deployment. The simulation framework provides a comprehensive and flexible methods to gain insights into COW-QKD approaches. The rest of this section first presents that simulation framework and describes the selected metrics for assessing the performance of the system.

\subsection{Simulation Framework}
The goal of the QKD simulations are to provide a practical and low-cost testing tool for extracting design guidelines before real-world deployment. The considered system model is presented~in~Fig.~\ref{Fig:qkd_arch}, from which it becomes evident that the Alice sends qubits through the quantum channel to the Bob. In turn, Bob measures the received qubits. In this process, the presence of an illegitimate receiver (Eve) affects the stochastic nature of the quantum channel and may lead to identification of eavesdropping. In more detail, COW depends on the inherent stochastic nature of quantum communications, in which information bits are encoded in superposition states of photons. Upon observation, the qubits have probabilistic measurement outcomes, which introduces fundamental randomness. Moreover, attenuation and noise, both of which are influenced by environmental factors, are introduced as quantum signals travel through the optical fiber, thus, increasing stochasticity and possibly introducing errors in the generated key. Security is enhanced by monitoring the interference of specific decoy pulses that are altered by eavesdropping attempts. All in all, these stochastic phenomena ensure the robustness and security of the COW protocol.

We assume that Alice generates $K$ pairs of entangled state as~in~\cite{Gao2022}
\begin{align}
        |Y\rangle & =\frac{1}{\sqrt{2}}\left(|+z\rangle_A\left|0_z\right\rangle_{A^{\prime}}+|-z\rangle_A\left|1_z\right\rangle_{A^{\prime}}\right),
\end{align}
or
\begin{align}
         |Y\rangle=\frac{\sqrt{N^{+}}}{2}|+x\rangle_A\left|0_x\right\rangle_{A^{\prime}}+\frac{\sqrt{N^{-}}}{2}|-x\rangle_A\left|1_x\right\rangle_{A^{\prime}},
\end{align}
with
\begin{align}
    &\left|0_z\right\rangle=|0\rangle_{2 n-1}|\alpha\rangle_{2 n}, \\
    &\left|1_z\right\rangle=|\alpha\rangle_{2 n-1}|0\rangle_{2 n}, \\
    &\left|0_x\right\rangle=\frac{\left|0_z\right\rangle+\left|1_z\right\rangle}{\sqrt{N^{+}}},
\end{align}
and 
\begin{align}
    \left|0_x\right\rangle=\frac{\left|0_z\right\rangle-\left|1_z\right\rangle}{\sqrt{N^{-}}} ,
\end{align}
denoting the $n$-th COW modes. In addition, 
\begin{align}
N^{ \pm}=2\left(1 \pm e^{-\mu}\right)
\end{align}
is the normalization factor, while $| \pm z\rangle$ and $| \pm x\rangle$ are eigenstates of the Pauli operators $Z$ and $X$, respectively. The subscript $A$ denotes the ancillary qubit held by Alice, and $A^{\prime}$ refers to the optical mode transmitted to Bob. Alice performs a random measurement of the ancillary qubit in either the $Z$ or $X$ basis, thereby obtaining the raw keys $\tilde{\mathbf{Z}}_A$ from the $Z$ basis and $\tilde{\mathbf{X}}_A$ from the $X$ basis. 

Subsequently, Alice sends the optical modes to Bob via an insecure quantum channel as illustrated in Fig.~\ref{Fig:qkd_arch}. The optical modes are detected either on the data line or the monitoring line. When a detector $D_T$ clicks at time $\mathcal{T}_0$ (or at a later time $\mathcal{T}_1$), Bob records the bit value $0$ (or $1$) as part of the raw key $\tilde{\mathbf{Z}}_B$. Additionally, Bob obtains the raw key $\tilde{\mathbf{X}}_B$ by observing the monitoring line, where a click of detector $D_{M_0}$ (or $D_{M_1}$) corresponds to a bit value of $0$ (or $1$). 

To characterize the quantum channel, we use its transmission $T\leq 1$ as well as its excess noise, i.e., $\nu$~\cite{Lodewyck2007}, and the path loss $l$. Thus, the noise variance at Bob can be expressed~as
\begin{align}
    N_{0,B} = \left(1+T\,\nu\right)\, N_0,
\end{align}
where $N_0$ stands for the Gaussian noise variance. The total channel added noise to the channel input can then be expressed~as
\begin{align}
    n_f = \frac{1}{T}-1 + \nu.
    \label{Eq:n_f}
\end{align}
In~\eqref{Eq:n_f}, $\frac{1}{T}-1$ represents the additional noise due to losses. By considering the homodyne detection, in the Bob detector, the introduced noise is given by
\begin{align}
    n = \frac{1+n_e}{\eta-1},
    \label{Eq:n}
\end{align}
where $n_e$ represents the thermal noise introduced by the receivers electronics, expressed in terms of shot noise. From~\eqref{Eq:n_f} and~\eqref{Eq:n}, the total noise added between Alice and Bob can be obtained~as
\begin{align}
    w = \frac{1}{T}-1 + \nu + \frac{1+n_e}{\left(\eta-1\right)\,T}. 
\end{align}

The path loss is defined as
\begin{align}
    l = 10^{0.1\,\kappa\,d},
\end{align}
where $\kappa$ stands for the fiber path-loss coefficient in $\rm{dB/km}$, and $d$ is the transmission distance.

\subsection{Performance Metrics}
The asymptotic secure key rate of the virtual entanglement-based protocol can be expressed as
\begin{align}
    \tilde{V}  =\frac{1}{\mathcal{P}_z K}\left[H_{\min }^\epsilon\left(\tilde{\mathbf{Z}}_A \mid E\right)-H_{\max}^\epsilon\left(\tilde{\mathbf{Z}}_A \mid \tilde{\mathbf{Z}}_B\right)\right] ,
    \label{Eq:Key_rate_0}
\end{align}
which can be equivalently written as~in~\cite{Tomamichel2011}
\begin{align}\
    \tilde{V} =\frac{1}{\mathcal{P}_z K}\left[n_z-H_{\max }^\epsilon\left(\tilde{\mathbf{X}}_A^{\prime} \mid B\right)-n_z f h\left(E_{\mathrm{z}}\right)\right] ,
    \label{Eq:Key_rate_1}
\end{align}
where $\mathcal{P}_z$ denotes the gain Alice measures in the $Z$ basis and Bob detects on the data line, $f$ represents the efficiency of error correction, and $E_z$ is the bit error rate in the $Z$ basis. In~\eqref{Eq:Key_rate}, $H_{\min }^\epsilon\left(\hat{\mathbf{Z}}_A \mid E\right)$ denotes the smooth min-entropy, which quantifies the average probability that Eve correctly guesses $\tilde{\mathbf{Z}}_A$ using the optimal strategy and her quantum memory correlations ~\cite{Konig2009}. Similarly, $H_{\text {max }}^\epsilon\left(\tilde{\mathbf{Z}}_A \mid \tilde{\mathbf{Z}}_B\right)$ represents the smooth max-entropy, indicating the number of additional bits needed to reconstruct $\tilde{\mathbf{Z}}_A$ from $\tilde{\mathbf{Z}}_B$ with a failure probability of $\epsilon$~\cite{Renes2012}. The bit string $\tilde{\mathrm{X}}_A^{\prime}$ represents what Alice would have obtained if she had measured in the $X$ basis, but was actually measured in the $Z$ basis. Consequently, the smooth max-entropy satisfies 
\begin{align}
    H_{\max }^\epsilon\left(\tilde{\mathbf{X}}_A^{\prime} \mid B\right) \leq n_z h\left(E_x\right) 
\end{align}
in the asymptotic limit, where $E_x$ denotes the bit error rate on the $X$ basis. 

The binary Shannon entropy is~written~as
\begin{align}
    h(a)=-a \log _2 a-(1-a) \log _2(1-a) ,
\end{align}
while the size of $\tilde{\mathbf{Z}}_A$ is denoted by $n_z$. 

Moreover,~\eqref{Eq:Key_rate_1} can be rewritten as
\begin{align}
    \tilde{V} =J_z\left[1-h\left(E_x\right)-f h\left(E_z\right)\right] ,
    \label{Eq:Key_rate}
\end{align}
in which 
\begin{align}
    J_z=\frac{n_z} {\mathcal{P}_z K}
 \end{align}
 or equivalently
 \begin{align}
J_z=\frac{1}{2} \left(J_{0_z}^{\mathcal{T}_0}+J_{0_z}^{\mathcal{T}_1}+J_{1_z}^{\mathcal{T}_0}+J_{1_z}^{\mathcal{T}_1}\right)
\end{align} 
is the likelihood Alice measures on the $Z$ basis and Bob detects on the data line.

The entropic uncertainty principle states that the signals detected by Bob's device are independent of Alice and Bob's basis choices~\cite{Tomamichel2012}. This requirement is naturally met when using active basis choice~\cite{Moroder2012, Wang2019}, but it needs to be carefully considered with passive basis choice. With passive basis choice, Eve can perform classical wavelength attacks~\cite{Li2011} to partially influence Bob's basis selection and induce weak basis-choice vulnerabilities. However, the secure key rate only decreases slightly when the wavelength is precisely characterized to minimize this effect~\cite{Li2015, Sun2020}. Notably in recent works, passive basis choice has been utilized extensively in a variety of QKD designs~\cite{Islam2017, Chen2021}. In this work, we assume that the beam-splitter is not under the control of the eavesdropper and employ the squashing model in the measurement setup~\cite{Beaudry2008, Cao2016}.

When Alice selects the $Z$ basis, she directly sends the optical modes $\left|0_z\right\rangle$ and $\left|1_z\right\rangle$ with a probability of $50\%$. Conversely, if Alice chooses the $X$ basis, she directly sends the COW4 optical modes $\left|0_x\right\rangle$ and $\left|1_x\right\rangle$ with probabilities $N^{+} / 4$ and $N^{-} / 4$, respectively. The density matrices of the Z and X bases, which are the same in COW4, can be~obtained~as
\begin{align}
        \rho & =\left(\left|0_z\right\rangle\left\langle 0_z|+| 1_z\right\rangle\left\langle 1_z\right|\right) / 2 
 \end{align}
 or equivalently
 \begin{align}
        \rho =\left(N^{+}\left|0_x\right\rangle\left\langle 0_x\left|+N^{-}\right| 1_x\right\rangle\left\langle 1_x\right|\right) / 4 .
\end{align}

Consequently, the QBER $E_z$ can be obtained from the measured gain as
\begin{align}
    E_z=\frac{J_{0_z}^{\mathcal{T}_1}+J_{1_z}^{\mathcal{T}_0}}{J_{0_z}^{\mathcal{T}_0}+J_{0_z}^{\mathcal{T}_1}+J_{1_z}^{\tau_0}+J_{1_z}^{\tau_1}} ,
\end{align}
where $J_{w_z}^{\mathcal{T}_j}$ denotes the gain for the event where Alice sends the optical mode $\left|w_z\right\rangle$ (with $w=0$ or 1) and Bob registers a click at the $\mathcal{T}_j$ moment (with $j=0$ or 1) on the data line. In addition, all possible events are bind by   \begin{align}
J_z=\frac{J_{0_z}^{\mathcal{T}_0}+J_{0_z}^{\mathcal{T}_1}+J_{1_z}^{\mathcal{T}_0}+J_{1_z}^{\mathcal{T}_1}}{2}.
\end{align}
Similarly, the bit error rate of the $X$ basis is~given~by
\begin{align}
    E_x =\frac{N^{+} J_{0_x}^{M_1}+\left[2\left(J_{0_z}^{M_0}+J_{1_z}^{M_0}\right)-N^{+} J_{0_x}^{M_0}\right]}{2\left(J_{0_z}^{M_0}+J_{0_z}^{M_1}+J_{1_z}^{M_0}+J_{1_z}^{M_1}\right)} ,
    \label{Eq:E_x}
\end{align}
where $J_{w_x(z)}^{M_i}$ represents the gain for the event where Alice sends the optical mode $\left|w_{x(z)}\right\rangle$ and Bob detects a click with detector $D_{M_i}$ on the monitoring line, with all event abiding~to 
\begin{align}
    N^{+} J_{0_x}^{M_i}+N^{-} J_{1_x}^{M_i}=2\left(J_{0_z}^{M_i}+J_{1_z}^{M_i}\right) .
\end{align}

Afterwards, Alice prepares an entangled state $|Y\rangle$ and measures the ancillary qubit in the $Z$ basis. Assuming that Alice prepares the encoding sequence $|0\rangle_{2 n-1}|\alpha\rangle_{2 n}$ or $|\alpha\rangle_{2 n-1}|0\rangle_{2 k}$ with equal probability, Eve cannot distinguish this procedure and, as a result, Alice's raw key is effectively identical to $\mathbf{Z}_A$. The secret key rate in the asymptotic limit can then be~expressed~as
\begin{align}
    V=J\left[1-h\left(E_{\mathrm{p}}^{\mathrm{u}}\right)-f h\left(E_{\mathrm{b}}\right)\right] ,
\end{align}
where $J=J_z$ and $E_b=E_z$ represent the gain and bit error rate, respectively. The phase error rate $E_{\mathrm{p}}^{\mathrm{u}}$ is the upper bound on the average error probability that Bob estimates as Alice's bit string, $\mathrm{X}_A^{\prime}$~\cite{Koashi2009}. This is equivalent to the upper bound on the bit error rate of the $X$ basis in COW4. In practice, since Alice does not measure qubits in the $X$ basis, the gains $J_{0_x}^{M_0}$ and $J_{0_x}^{M_1}$ cannot be directly obtained in COW4. However, we can use the gains of other quantum states, such as $|\alpha\rangle_{2 n-1}|\alpha\rangle_{2 n}$ and $|0\rangle_{2 n-1}|0\rangle_{2 n}$, which are directly measurable in the COW-QKD protocol, to estimate the upper bound, $\bar{J}_{0_x}^{M_1}$, and the lower bound, $\bar{J}_{0_x}^{M_0}$. In this sense, the phase error rate $E_{\mathrm{p}}^{\mathrm{u}}$ can be~expressed~as
\begin{align}
    E_{\mathrm{p}}^{\mathrm{u}}=\frac{N^{+} \bar{J}_{0_x}^{M_1}+\left[2\left(J_{0_z}^{M_0}+J_{1_z}^{M_0}\right)-N^{+} \bar{J}_{0_x}^{M_0}\right]}{2\left(J_{0_z}^{M_0}+J_{0_z}^{M_1}+J_{1_z}^{M_0}+J_{1_z}^{M_1}\right)}
\end{align}
with $J_{0_z}^{M_i}=J_{0 \alpha}^{M_i}$ and $J_{1_z}^{M_i}=J_{\alpha 0}^{M_i}$. Under collective attacks, the upper $\bar{J}_{0_x}^{M_1}$ and the lower bound $\bar{J}_{0_x}^{M_0}$ can be obtained as~in~\cite{curty2018, Wang2019}.

\section{QKD Implementation} \label{sec:experimental-setup}

\begin{figure}
    \centering
    \includegraphics[width=1\linewidth]{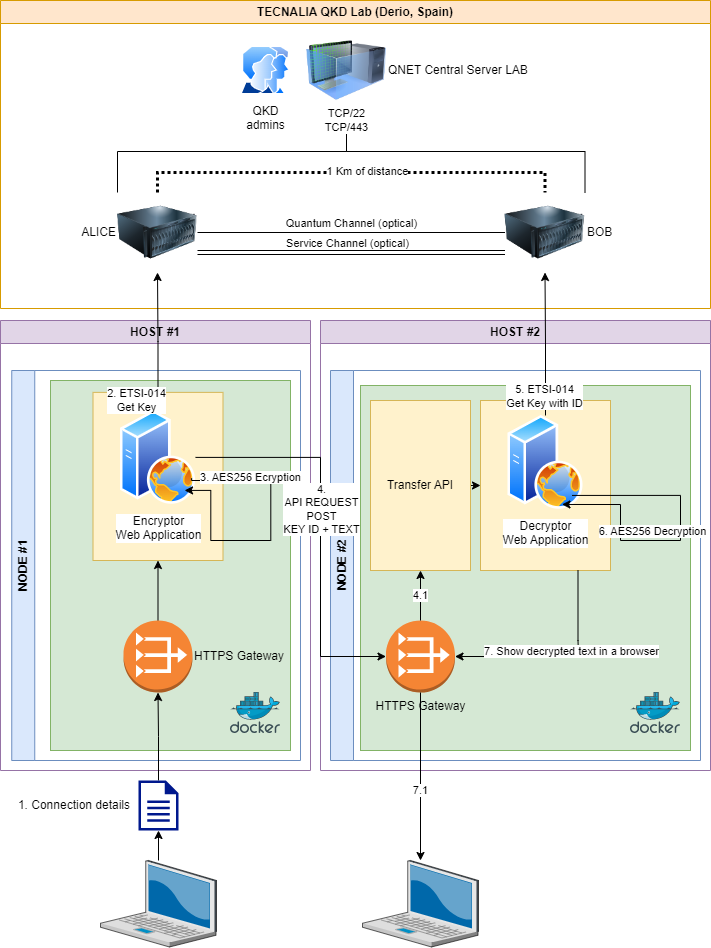}
    \caption{QKD Applications architecture.}
    \label{fig:demo-apps-architecture}
\end{figure}

Figure~\ref{fig:demo-apps-architecture} provides the architecture of the QKD expiremental implementation. Two hosts are deployed in different locations. The first host (Host 1) is set to ALICE's node, while the second to Bob's node (Host 2). Both hosts are containers and have been developed using the standard interfaces of QKD devices. Structuring the nodes as independent applications allows us to adopt the ETSI standard QKD application programming interfaces (API) as well as to enable rapid deployment and platform independence during the experiments. To interact with the QKD REST API, each application possesses a pre-deployed client certificate, which is used to authenticate and retrieve cryptographic keys from either the Alice or Bob side. These client certificates are signed by a self-signed authority located on the QKD devices and are deployed within Docker containers prior to the start of communication. The applications have been developed using a robust software stack where the backend is built with Rust as the programming language~\cite{Rust}, Actix as the web framework~\cite{Actix}, and PostgreSQL as the database~\cite{PostgreSQL}. The front-end is implemented using WebAssembly as the programming language~\cite{WebAssembly}, Trunk as the asset bundler~\cite{Trunk}, YEW-RS as the web framework~\cite{YEW-RS}, and TailwindCSS for styling~\cite{TailwindCSS}.

We assume that the communication is initiated by Bob, who  requests a key from Alice through the QKD channel. Once the key is received, Bob encrypts the data.  Next, he sends the encrypted data along with the key identifier to Alice. Upon receiving the information, Alice requests a key from Bob through the QKD equipment at his end. Using the key provided by Bob, Alice decrypts the received data. This signifies that the communication between the two nodes is established securely.

The quantum channel of the QKD devices is a dark fiber where the devices exchange quantum states on photons with extreme sensitivity to medium imperfections, is employed. Two other dark fibers, which form the classic channel, are needed to perform synchronization tasks to support the quantum channel information exchange. The service layer contains the SAEs, which do not directly interact with either the quantum or classical channels. To manage the cryptographic material generated by QKD and store a buffer to be consumed by classical applications as a service, the QKD devices are packed with an integrated KMS. This software layer is an intermediate between the key consuming applications and the quantum information exchange. 

Since 2010, the ETSI has issued a set of standards for QKD vendors to unify all interfaces related to this technology. Among others, there are two specifying how an SAE can interact with the classical KMS layer to request a key of a specific length:
\begin{itemize}
    \item ETSI GS QKD 004: Quantum key distribution application interface (API)
    \item ETSI GS QKD 014: Quantum key distribution protocol and data format of REST-based key delivery API
\end{itemize}
Both interfaces serve the same goal. The main difference is regarding the implementation details. Since ETSI-004 is implementation agnostic and provides the concept of sessions, ETSI-014 is specified as an REST API with Mutual Authentication enabled. In many cases, choosing one or another depends on which is supported by the QKD device vendor. In this case, the equipment used for the experiments is only compatible with ETSI-014. 

The complete specification of ETSI-014 API can be accessed publicly in the ETSI repository~\cite{ETSI014}. The interface is enforced with mutual TLS to protect the KMS from unauthorized users. This means that all SAEs, such each BS, must own a client certificate issued by the public key Infrastructure hosted at QKD modules. These certificates have been previously deployed as personal information exchange (PFX) files within the source files of each application in the Docker containers. In addition to the quantum and classical channels, each QKD module must be available via local area network (LAN) with the SAEs for which it provides service. Each application Docker runs in a machine on the same internet protocol (IP) subnet as each QKD module.

\subsection{Methodology}
To systematically stress test the experimentation testbed, an automated script was developed to reproduce a continuous communication flow with varying time intervals and total durations. This script functions as an automation layer, managing the application functionality through its REST API, a process that could alternatively be executed manually via the graphical user interface. Furthermore, the script executes Algorithm~\ref{alg:encryption-process}, ensuring an efficient and repeatable testing process.

\begin{algorithm}
\caption{Connection data payload encryption and decryption}
\label{alg:encryption-process}
\begin{algorithmic}[1]
\For{$N$ iterations}
    \State Set an example connection data payload to be securely sent
    \State Request NODE1 API to encrypt the payload
    \State \textbf{NODE1:}
    \State \quad Request Alice for an encryption key over the conventional channel 
    \State \quad Encrypt the payload with the key using AES, 3DES, or others
    \State \quad Return to the client: \textit{KeyID}, description, ciphertext, KMS response time
    \State Request NODE2 API to add the encrypted connection data to its database
    \State Request NODE2 API to decrypt the previously sent data
    \State \textbf{NODE2:}
    \State \quad Fetch the connection data from the database
    \State \quad Request Bob for the decryption key using the associated \textit{KeyID}
    \State \quad Decrypt the payload with the key using AES256 and store the plaintext in the database
    \State \quad Return to the client: status, KMS response time
    \State Request NODE2 API to fetch the decrypted record
    \State \quad Fetch the record from the database
    \State \quad Return to the client all record columns
    \State Show the decrypted connection data on the terminal
    \State Store the timestamp and the KMS response time in a log file for analysis
    \State Wait $s$ seconds
\EndFor
\end{algorithmic}
\end{algorithm}

The script was used as the main tool to operate the infrastructure during the experiments. The duration and the key length were set to $1$ hour and $256$ bits, respectively. The collected metrics include the quantum bit error rate (QBER) and the secrecy key rate (KeyRate). The experimental parameters of the experiments are illustrated~in~Table~\ref{Tab:Experimental_parameters}.%

\begin{table}
    \caption{Experimental parameters}
    \centering
    \begin{tabular}{r|l}
        Parameter & Value \\
        \hline
        \hline
        Amplifier voltage & $8.46\text{ V}$ \\
        DC voltage & $6.6\text{ V}$ \\
        Laser bias & $50\text{ A}$ \\
        Laser power & $654\text{ $\mu$W}$ \\
        PD dead time & $50\text{ $\mu$s}$ \\
        Pulse width & $600\text{ ns}$ \\
        QBER integration time & $10\text{ s}$
    \end{tabular}
    \label{Tab:Experimental_parameters}
\end{table}

\subsection{Test Scenarios \& Deployment}
The experiments will be conducted across various deployment scenarios to identify performance deviations in specific setups and protocols supported by QKD devices. The infrastructure will include actual commercial QKD equipment and a simulator. The QKD protocols to be analyzed are COW with 3 and 4 states. Fibre deployments will involve dark fibre setups at different distances: 1 km (actual deployment in Tecnalia premises), 2 km (fibre launch lead in the laboratory), and 3.5 km (a combination of 1.5 km and 2 km fibre launch leads in the laboratory). Operations will be tested under normal conditions and in the presence of eavesdropping.

\begin{figure}[!ht]
    \centering
    \includegraphics[width=1\linewidth]{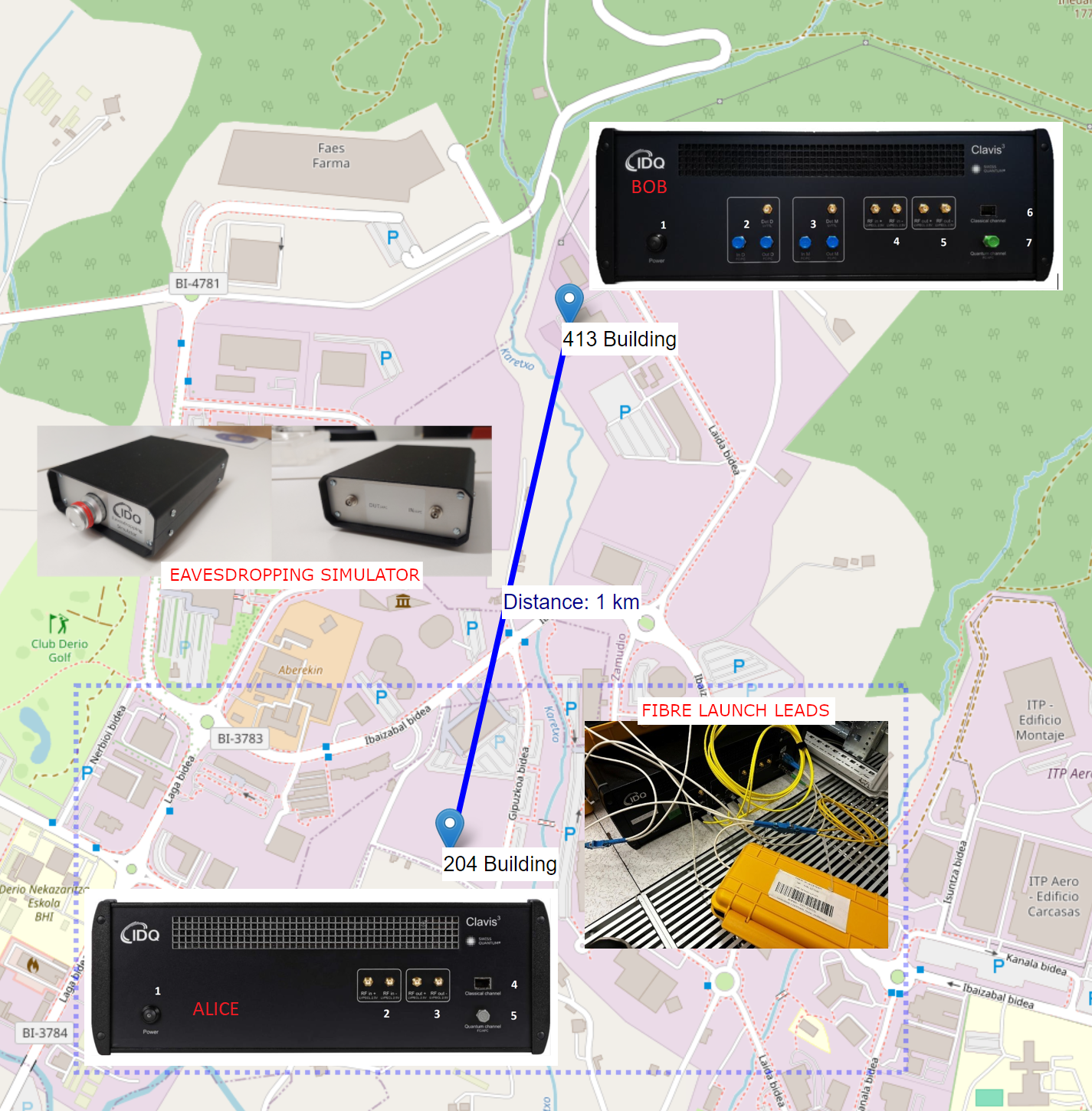}
    \caption{QKD fibre link and QKD equipment.}
    \label{fig:qkd-map}
\end{figure}

The commercial equipment used for the experiments is the Clavis3 QKD Platform made by IDQuantique~\cite{IDQuantique}. The equipment is deployed in the server rooms of two separate buildings of TECNALIA premises in Derio (Spain). The municipality owns the fibre deployment, and the only requirement was three dark fibres without any active switching to avoid affecting the quantum states. The map in Fig.~\ref{fig:qkd-map} illustrates the dark fibre link, with the blue line representing the approximate fibre optics connection between the two buildings. To simulate extended distances, two fibre launch leads, measuring 1.5 km and 2 km respectively, were deployed in the laboratory. Both leads were connected through an optical splitter to achieve the maximum length, as depicted in the figure.

This experiment setup is unique compared to previous studies due to several key differences. It integrates the entire encryption service stack, building a bespoke application on top of the QKD service to replicate, as closely as possible, the real-world use of this technology. This approach enables the capture of dynamic behaviours in service performance, influenced by components such as the key management system's throughput or the classical network required to connect each node. Additionally, unlike prior experiments that focused solely on the quantum channel or discrete variable protocols, such as the BB84, our study examines a more realistic deployment of a QKD link with off-the-self COW equipment. It incorporates a protocol commonly used in commercial equipment, offering a more comprehensive understanding of its performance in real scenarios and demonstrating how an application using the existing ETSI QKD standards can benefit from the cryptographic material generated by the quantum link. Furthermore, the utilisation of previously deployed dark fibres, installed over a decade ago by the municipality without QKD technology in mind, distinguishes our setup. This scenario closely mirrors many potential QKD deployments by companies or public organisations, where deploying new infrastructure is challenging, and existing resources must be optimised. These unique aspects ensure that our experiment addresses gaps in the literature while providing novel and practical insights.

The equipment reports parameter statistics in real time to a server running quantum management system (QMS), a monitoring software. This server is a virtual machine running the software in Docker containers. The web application run by the QMS provides a graphic user interface to set up the platform and visualize all the parameters measured by the systems (e.g., QBER and KeyRate). This tool is essential for obtaining the measurements after the experimentation ends. The QMS only has a connectivity requirement, which is to have IP communication with each device. As both Alice and Bob have been deployed in TECNALIA premises both are connected by ethernet to a subnet open to the QMS server. A graphical representation of the network can be found in Fig.~\ref{fig:demo-apps-architecture}. To perform the eavesdropping tests the vendor IDQuantique provides an additional tool for this purpose. The eavesdropping simulator hardware is presented in~Fig.~\ref{fig:qkd-map}.

\section{Results \& Discussions}
\label{sec:results-and-discussion}
The following section presents a detailed analysis of the COW QKD protocol performance as observed in our simulation and benchmarked against data from experimental setups. By leveraging both simulated and real data, we gain insights into critical metrics such as key generation rate, QBER, and protocol robustness under varying conditions.

\subsection{Experiments vs Simulations}
\label{sec:exp-vs-sim}
This section presents comparative results between the experiments and the simulations for the COW3 and COW4 protocols. These result not only verify the validity of the simulations but also highlight the capabilities of the simulation.

\begin{figure}
    \centering
    \begin{subfigure}[b]{0.48\textwidth}
        \centering
        \includegraphics[width=0.9\linewidth]{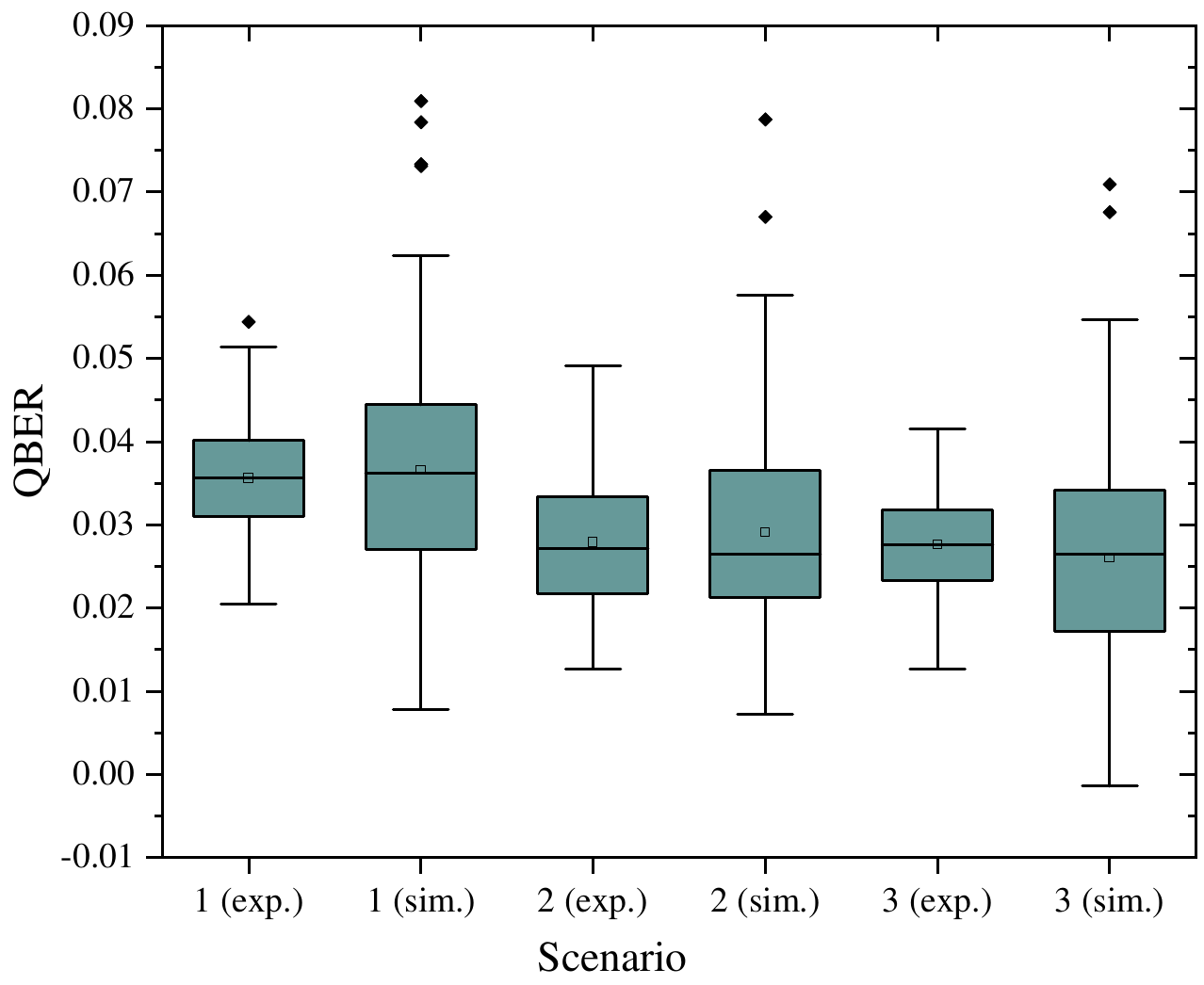}
        \caption{}
        \label{Fig:cow3_qber}
    \end{subfigure}
    \hfill
    \begin{subfigure}[b]{0.48\textwidth}
        \centering
        \includegraphics[width=0.9\linewidth]{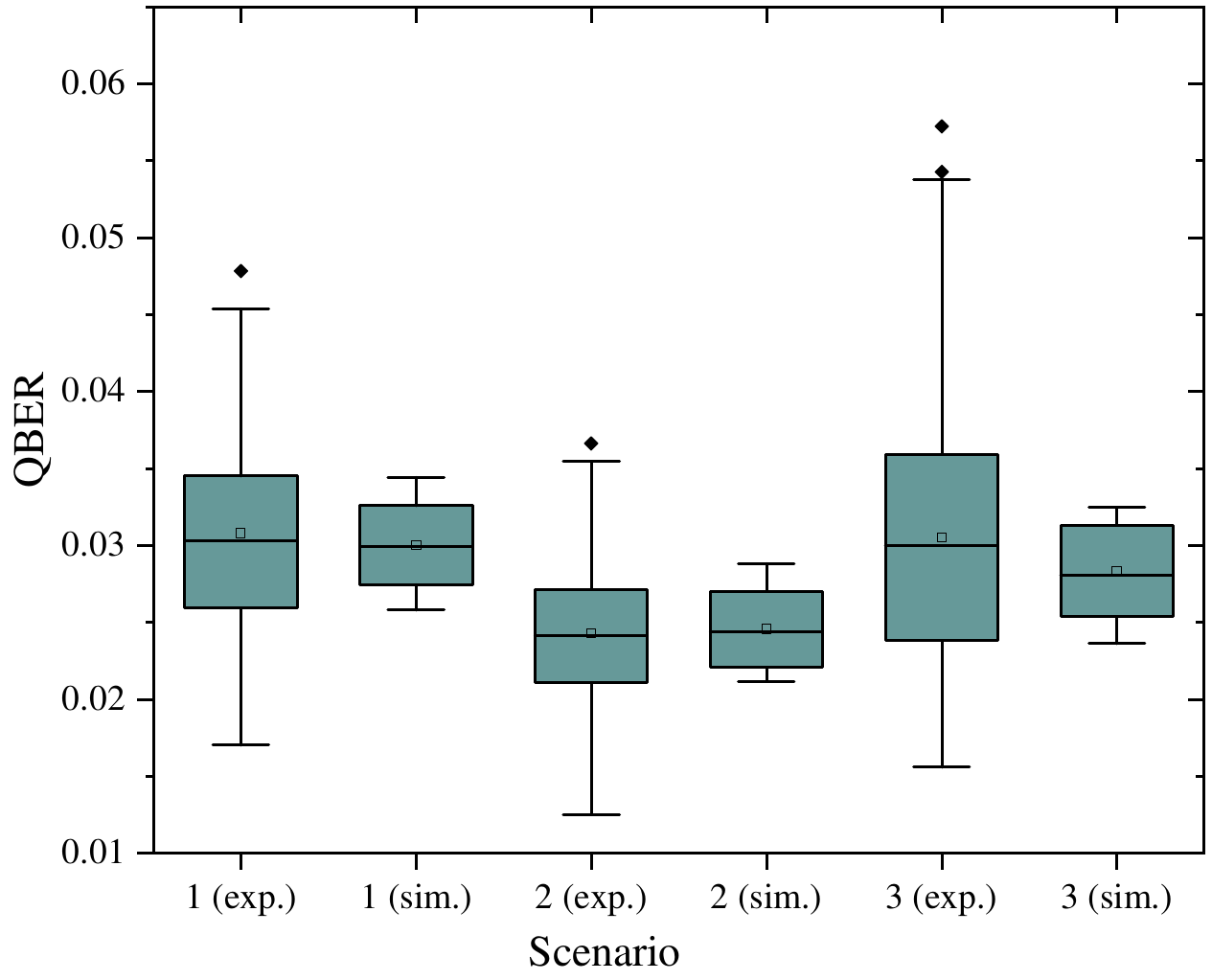}
        \caption{}
        \label{Fig:cow4_qber}
    \end{subfigure}
    \caption{QBER of a) COW3 and b) COW4.}
\end{figure}

In Figs.~\ref{Fig:cow3_qber} and~\ref{Fig:cow4_qber}, the QBER is evaluated in three scenarios of experimental and simulated data based on different fiber lengths, i.e., $1$, $2$, and $3.5$ km. Across these distances, both protocols follow similar patterns that showcase an average QBER of about $0.03$, thus, illustrating stable performance irrespective of the fiber length. The collected experimental results is further validated by other works that have observed similar results~\cite{TelloCastillo,Stucki2005,Stucki_2009,Stucki_2009b}. While both experiments align with the simulated data in terms of the average QBER, they are characterized by pronounced spikes, likely due to external factors. This is a result of the stability of the controlled simulation environment. Finally, COW4 has smaller deviations around its average QBER value as compared to COW3, suggesting that COW4 could be more stable over time, particularly for greater lengths of fiber.

\begin{figure}
    \centering
    \begin{subfigure}[b]{0.48\textwidth}
        \centering
        \includegraphics[width=0.9\linewidth]{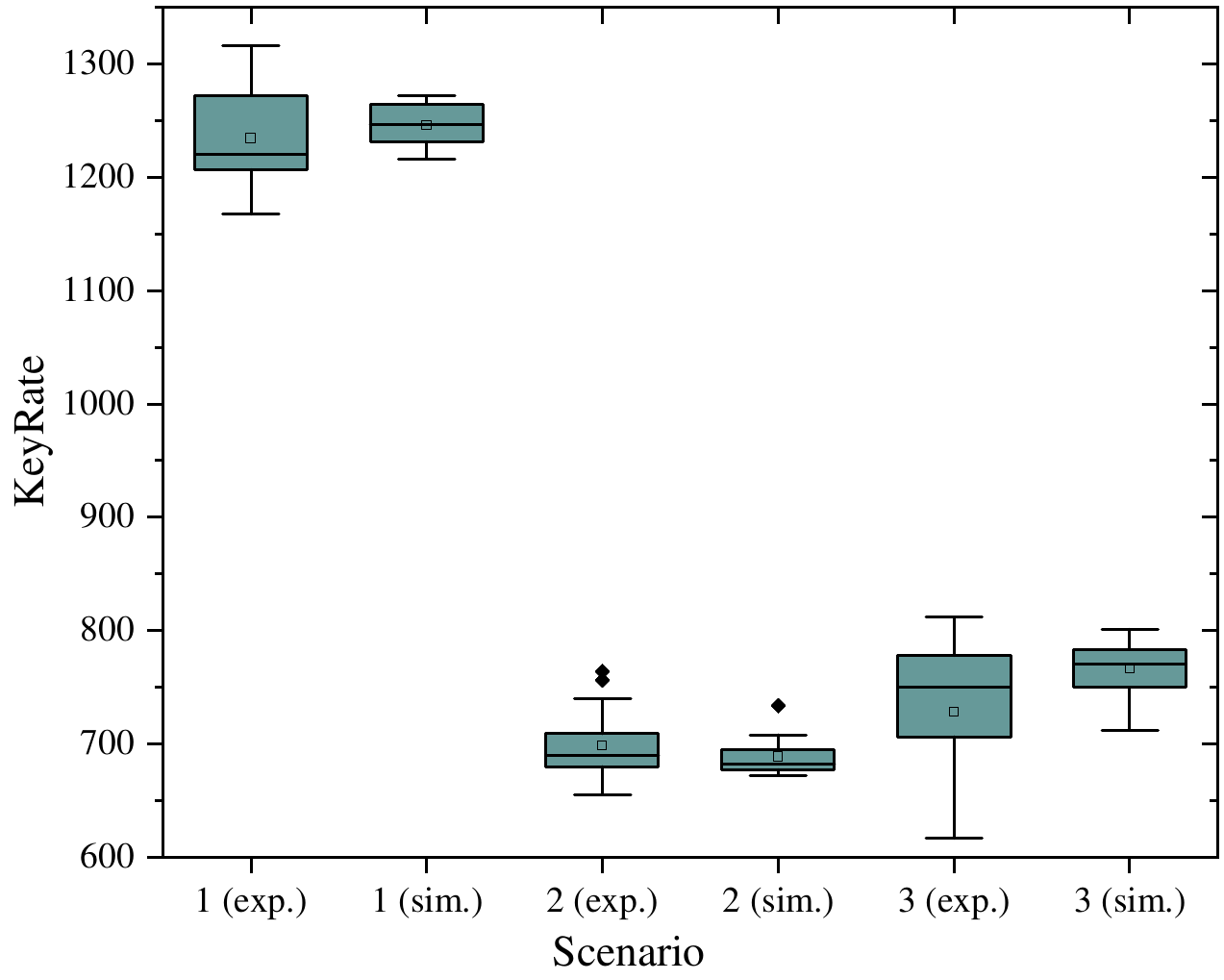}
        \caption{}
        \label{Fig:cow3_keyrate}
    \end{subfigure}
    \hfill
    \begin{subfigure}[b]{0.48\textwidth}
        \centering
        \includegraphics[width=0.9\linewidth]{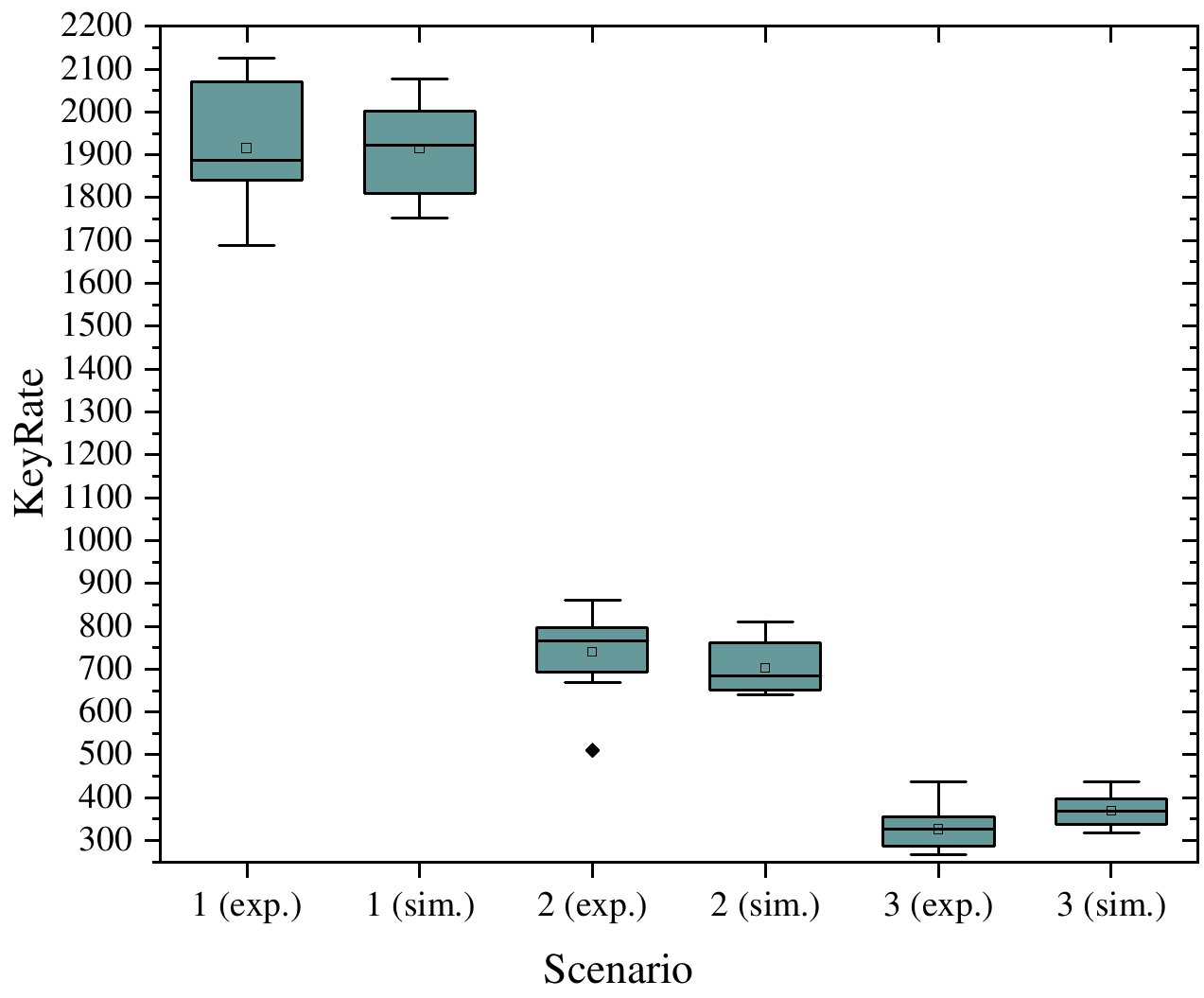}
        \caption{}
        \label{Fig:cow4_keyrate}
    \end{subfigure}
    \caption{Keyrate of a) COW3 and b) COW4.}
\end{figure}

Figures~\ref{Fig:cow3_keyrate}~and~\ref{Fig:cow4_keyrate} present the KeyRate as a function of time for the COW3 and COW4 protocols. For each figure, experimental and simulated results are presented from the aforementioned three scenarios ($1$, $2$, and $3.5$ km). For COW4, the KeyRate attained by the simulations and the experiments are very close for all three scenarios. Specifically, COW4 achieves a KeyRate of approximately $2000$, $700$, and $400$ bits/s in the $1$, $2$, and $3.5$ km scenarios, respectively. This verifies the validity of the simulation framework. In COW3, for $1$ km, the experiments showcased an average of $1250$ bits/sec, while the simulations exhibit $1150$ bits/sec. For $2$ km, the experimental KeyRate is about $850$ bits/sec and the simulation one is $750$ bits/sec. Also, for $3.5$ km, the experimental and simulation KeyRate are equal to $700$ and $650$ bits/sec, respectively. It becomes evident that in all scenarios the experimental results achieve a slightly higher KeyRate that the simulated ones. This is because the commercial hardware that was utilized for realizing the experiments tries to dynamically optimize the transmission and, therefore, attains higher performance. Finally, this is a strong point of the simulator, since its output is slightly more pessimistic than the achievable one, which provide more confidence in the applicability of its estimations. In other words, for COW3, simulations can be used as lower bounds.

\subsection{Design Guidelines}
\label{sec:guidelines}
This section presents simulation results that illustrate the impact of various QKD design parameters on the overall system's performance. Insightful discussions are also provided in order to extract design guidelines for QKD systems even before their deployment.

\begin{figure}
    \centering
    \includegraphics[width=0.9\linewidth]{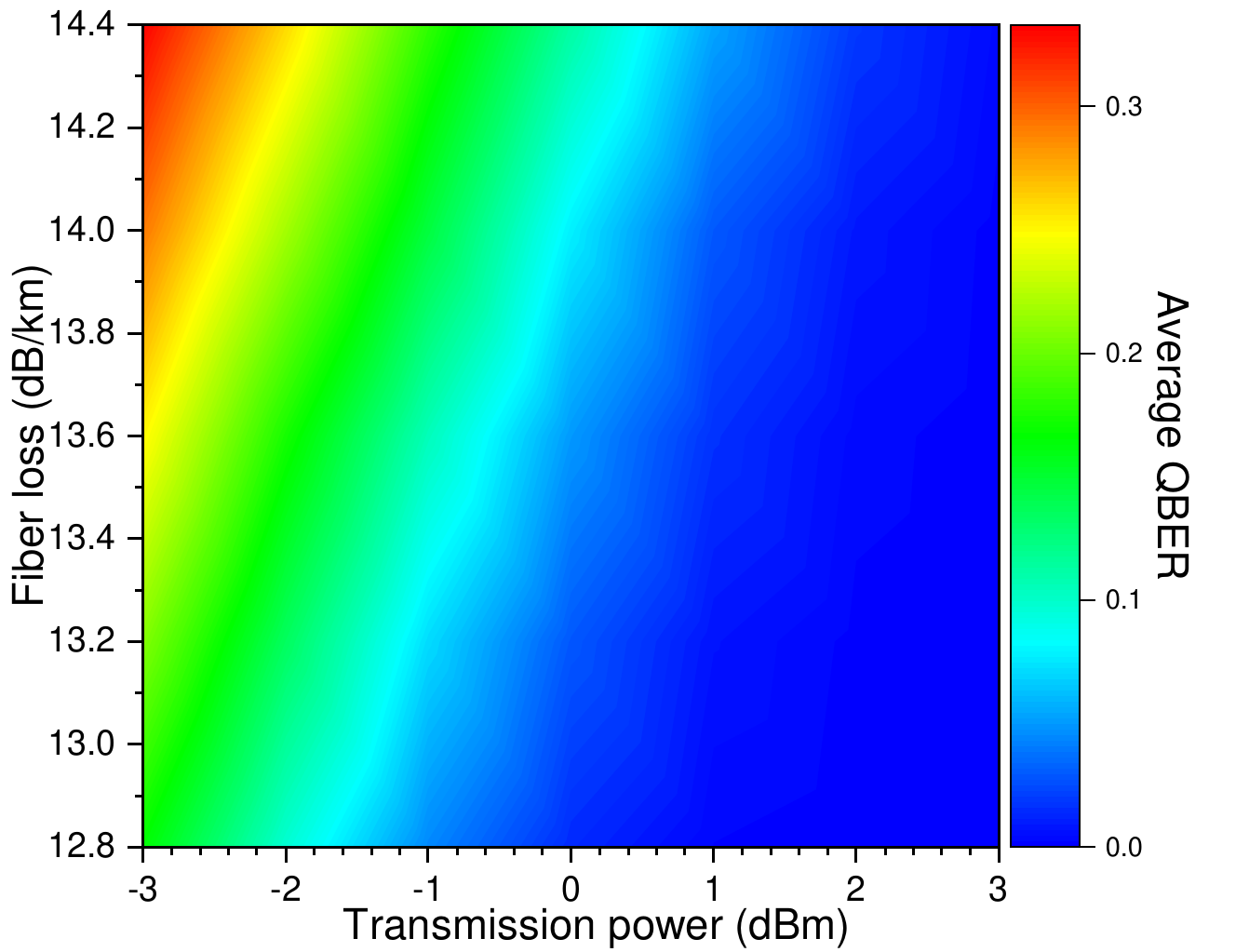}
    \caption{Average QBER as a function of transmission power and fiber loss.}
    \label{Fig:qber_loss_power}
\end{figure}
Figure~\ref{Fig:qber_loss_power} illustrates the link quality between fiber loss and transmission power in terms of average QBER. The data highlight that increasing transmission power results in poorer QBER values regardless of whether the fiber loss is modest or high. In contrast, reduced transmission power is related with higher average QBER values, with the highest QBER recorded when fiber loss is at its maximum. Furthermore, the plot displays a distinct zone for transmission power higher than 1 dBm in which the average QBER is continuously low, regardless of fiber loss value. This suggests a threshold, after which power above a certain level reduces the impact of fiber loss on QBER. According to this figure, increasing transmission power is a successful method for reducing QBER in a communication system, even in cases where fiber loss is high. The region where power surpasses 1 dBm indicates a minimum power threshold that must be met in order to keep QBER low for a fiber that is characterized by 14.4 dB/km loss. If we substitute this fiber with another of a lower loss link, it is possible to attain the same QBER with lower transmission power. Therefore, for best performance and minimal error rates, maintaining transmission power over a certain level depending on the fiber losses is critical.

\begin{figure}
    \centering
    \includegraphics[width=0.9\linewidth]{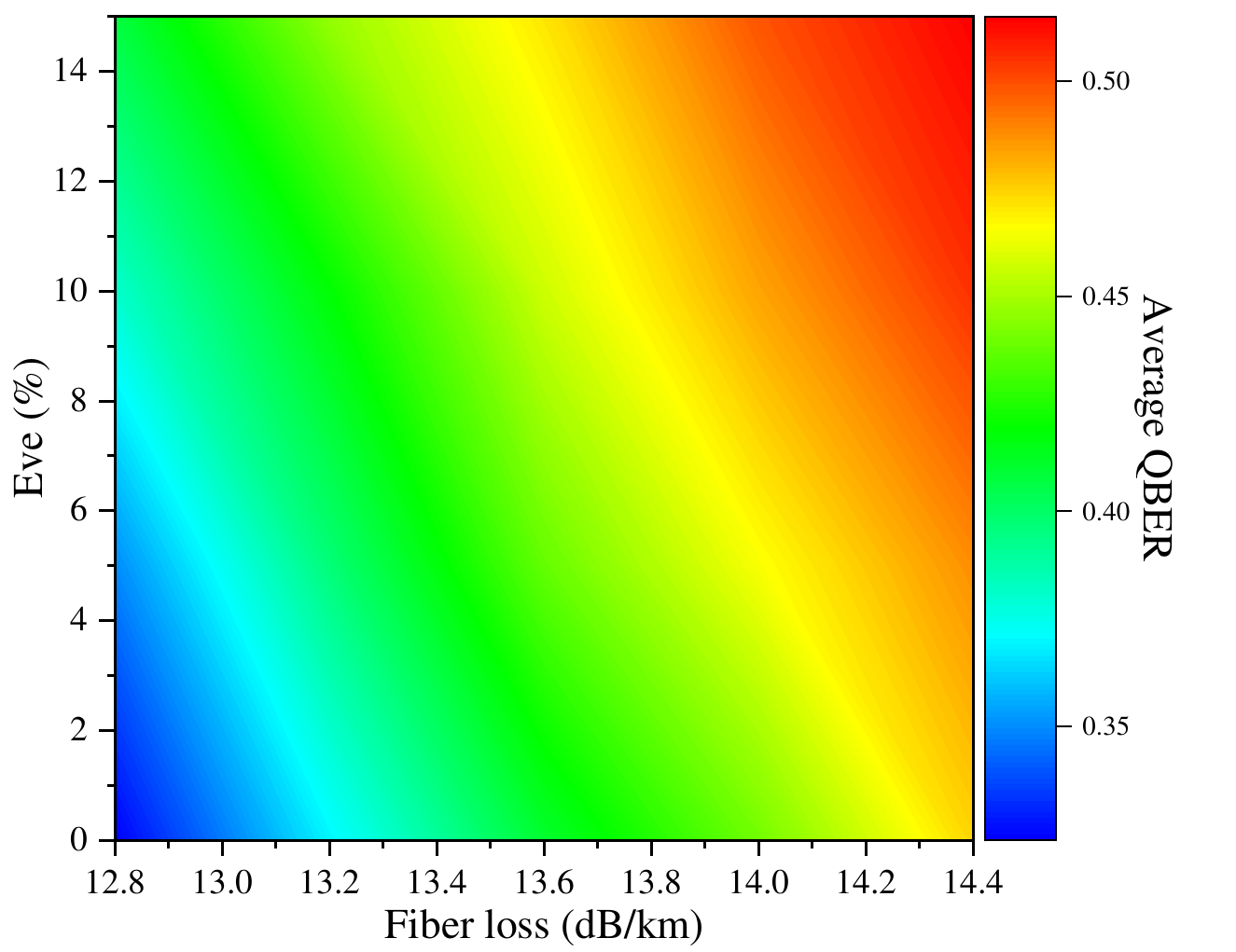}
    \caption{Average QBER as a function of fiber loss for different eavesdropping scenarios.}
    \label{Fig:qber_loss_eve}
\end{figure}
Figure~\ref{Fig:qber_loss_eve} depicts the average QBER for different eavesdropping and fiber loss configurations. As expected, there is a clear relationship between the eavesdropping scale and the average QBER. Specifically, higher eavesdropping values result in higher QBER. When both the eavesdropping percentage and fiber loss are at their peak, QBER attains the highest values as well. For the best fiber loss configuration, the average QBER stays low, even if the scale of eavesdropping is significant. In contrast, at large levels of fiber loss, the QBER rises significantly independent of the eavesdropping scale, demonstrating that fiber loss can have a major impact on the system. This suggests that maintaining low fiber loss is critical for keeping low error rates, even in the face of eavesdropping. Reducing fiber loss should be a top goal in communication system security since it lowers the possibility of QBER increase, even in the presence of eavesdropping. All in all, we notice that both eavesdropping scale and fiber loss alone lead to a rise in QBER, with their combined effects being extremely damaging to the system's security.  

\begin{figure}
    \centering
    \includegraphics[width=0.9\linewidth]{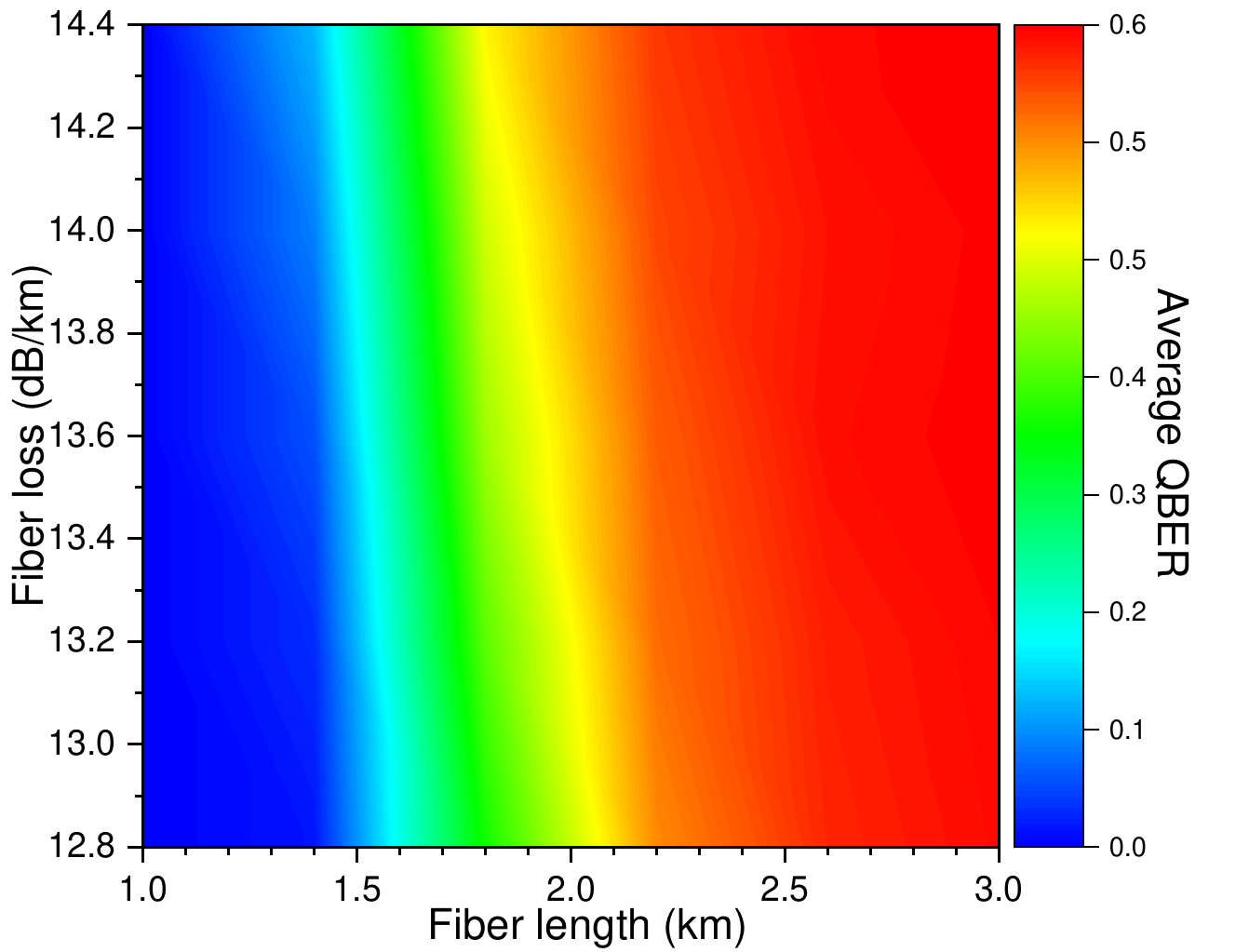}
    \caption{Average QBER as a function of fiber lenght and loss.}
    \label{Fig:qber_km_loss}
\end{figure}
Figure~\ref{Fig:qber_km_loss} shows the link quality between fiber loss and fiber length measured through the average QBER. As expected, as the transmission distance increases, the average QBER increases as well. It is also worth noting that a minor initial increase in fiber length causes a noticeable spike in QBER, which then varies directly with subsequent changes in fiber length across the whole spectrum of fiber loss. Moreover, even a small increase in fiber length can cause a notable change in QBER, thereby underlining the need of precisely regulating and minimizing fiber length to keep low error rates in communication systems. 

\begin{figure}
    \centering
    \includegraphics[width=0.9\linewidth]{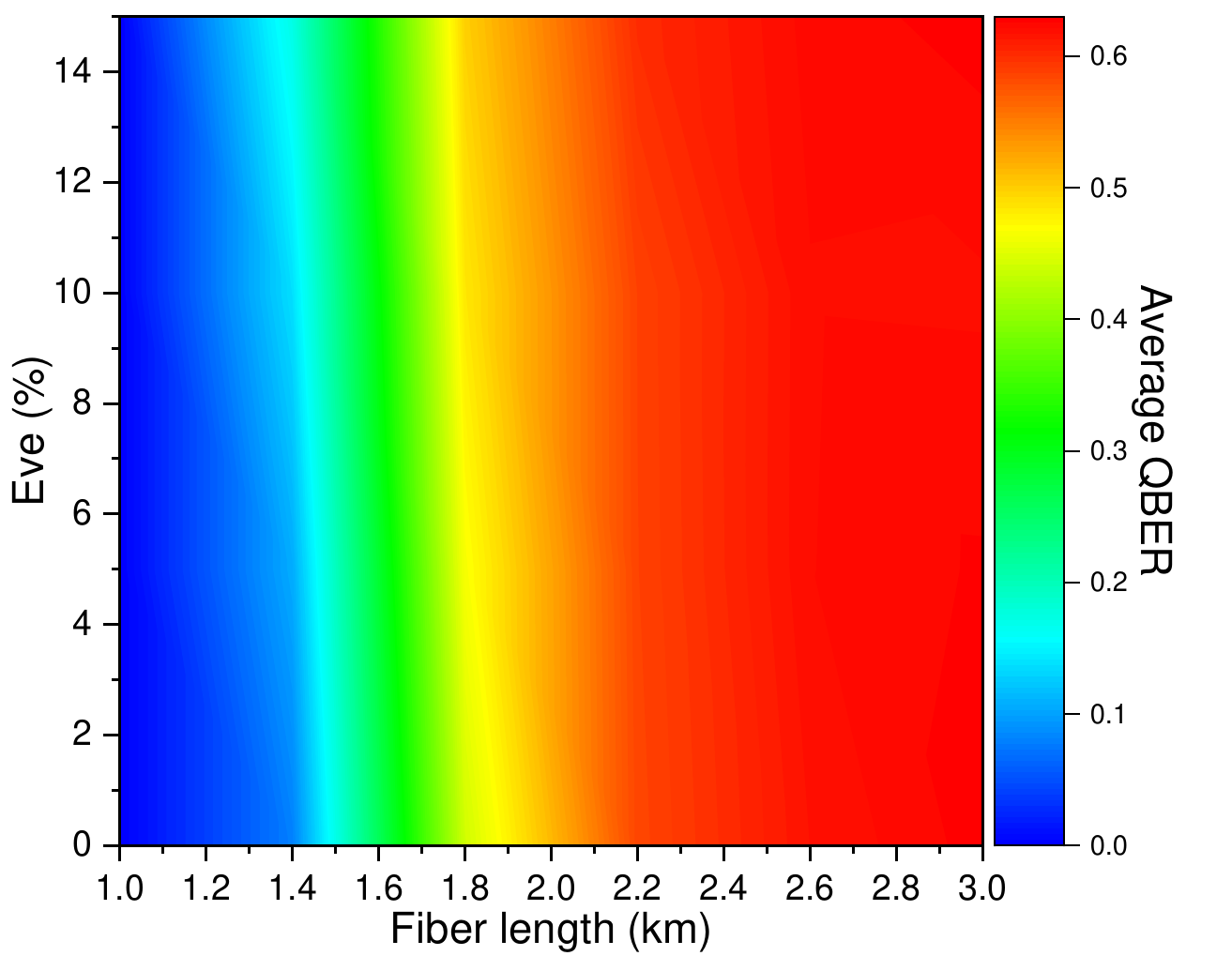}
    \caption{Average QBER as a function of fiber length for different eavesdropping scenarios.}
    \label{Fig:qber_km_eve}
\end{figure}
In Fig.~\ref{Fig:qber_km_eve} the average QBER is shown as a function of the eavesdropping scale and fiber length. The results show that the shortest fiber lengths have the lowest average QBER values, independent of eavesdropping level. Once the fiber length exceeds a threshold of around one-third of its total values, the QBER achieves its maximal value, indicating a critical limit beyond which further increases in fiber length have no effect on error rates. On the other hand, the average QBER remains consistent across the whole range of eavesdropping scales. This illustrates that fiber length is a dominant factor in determining average QBER, outweighing the impacts of the eavesdropping scale.  

\begin{figure}
    \centering
    \includegraphics[width=0.9\linewidth]{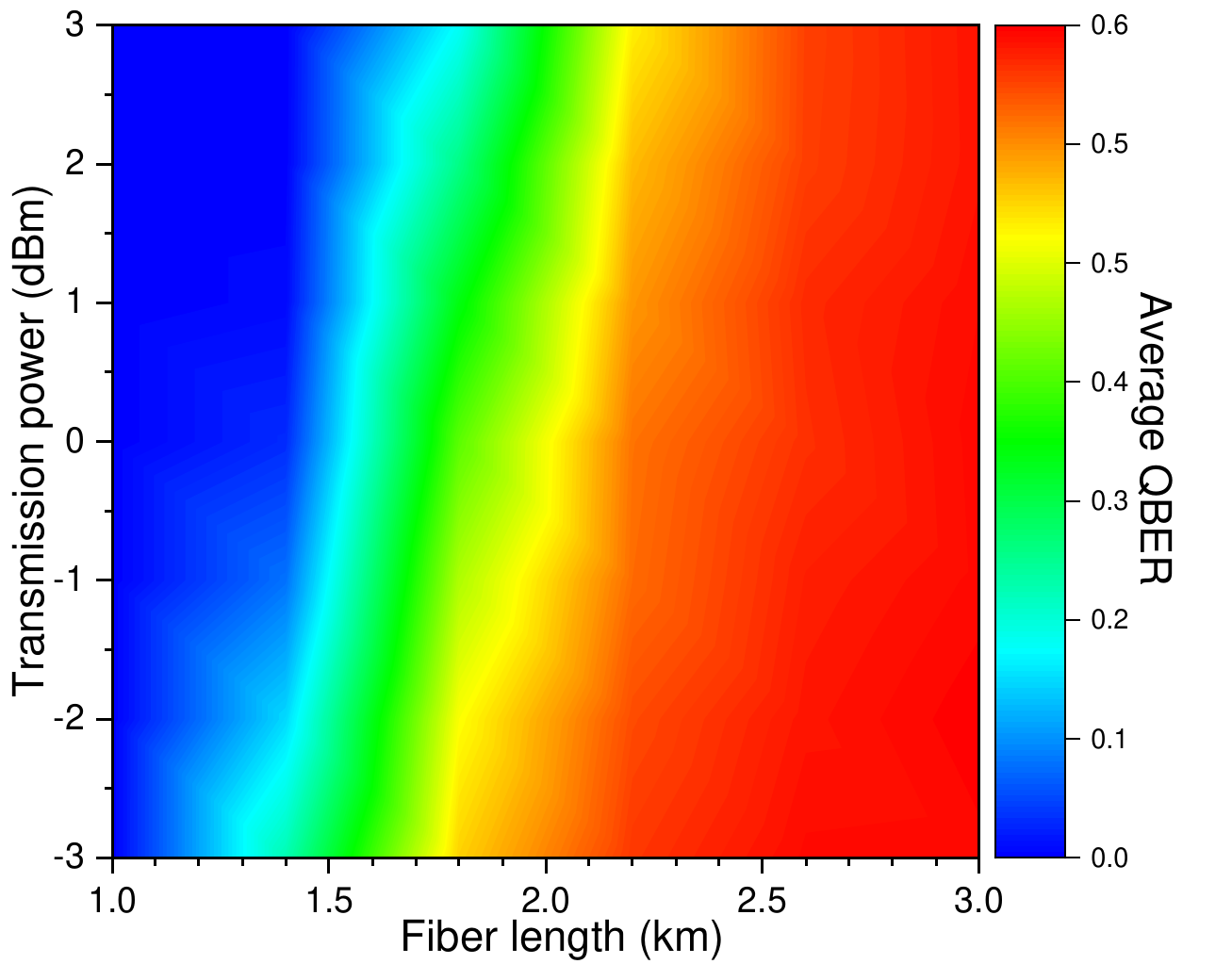}
    \caption{Average QBER as a function of transmission power and fiber length.}
    \label{Fig:qber_km_power}
\end{figure}
Fig.~\ref{Fig:qber_km_power} presents the average QBER for different values of transmission power and fiber. As anticipated, the figure indicates that the longest fiber lengths correlate with the highest average QBER values. Notably, over a particular fiber length threshold, the average QBER achieves its peak value regardless of the transmission power level. In contrast, at the shortest fiber lengths, the curve exhibits the lowest QBER values, regardless of power levels. This shows that reducing fiber length is critical for maintaining low QBER. Fiber length plays an important role in determining the average QBER in a communication system. While increasing transmission power can generally aid in reducing QBER, this effect fades once a particular fiber length threshold is reached, after which QBER remains high independent of power modifications. To maintain best performance and decrease error rates, fiber lengths should be kept to the shortest possible value. Finally, we emphasize that, in terms of preserving low QBER levels over extended transmission distances, measures aimed at reducing or optimizing fiber length may be more beneficial than just boosting transmission power.

\section{Conclusions}
\label{sec:conclusion}
This paper has presented an experimental-verified QKD simulation framework that accounts for the fibers and end-devices characteristics. In particular, we have documented the key operation elements of the QKD and reported the feasibility of COW QKD solutions.  Next, we have presented the corresponding simulation framework as well as the definition of the key performance metrics. To verify the accuracy of the proposed simulation framework, we have deployed a real-world QKD system. A number of experiments have been  performed for diverse deployments in the presence or absence of an eavesdropper. Simulations and experimental results have shown an acceptable match; a fact that renders our proposed simulation framework a suitable tool for designing QKD systems.  
\balance
\bibliographystyle{IEEEtran}
\bibliography{references}

\begin{thebibliography}{10}
\providecommand{\url}[1]{#1}
\csname url@samestyle\endcsname
\providecommand{\newblock}{\relax}
\providecommand{\bibinfo}[2]{#2}
\providecommand{\BIBentrySTDinterwordspacing}{\spaceskip=0pt\relax}
\providecommand{\BIBentryALTinterwordstretchfactor}{4}
\providecommand{\BIBentryALTinterwordspacing}{\spaceskip=\fontdimen2\font plus
\BIBentryALTinterwordstretchfactor\fontdimen3\font minus \fontdimen4\font\relax}
\providecommand{\BIBforeignlanguage}[2]{{%
\expandafter\ifx\csname l@#1\endcsname\relax
\typeout{** WARNING: IEEEtran.bst: No hyphenation pattern has been}%
\typeout{** loaded for the language `#1'. Using the pattern for}%
\typeout{** the default language instead.}%
\else
\language=\csname l@#1\endcsname
\fi
#2}}
\providecommand{\BIBdecl}{\relax}
\BIBdecl

\bibitem{Je2021}
D.~Je, J.~Jung, and S.~Choi, ``Toward 6g security: Technology trends, threats, and solutions,'' \emph{IEEE Communications Standards Magazine}, vol.~5, no.~3, pp. 64--71, 2021.

\bibitem{atutxa24}
A.~Atutxa \emph{et~al.}, ``{Towards a quantum-safe 5G: Quantum Key Distribution in core networks},'' \emph{Computer Communications}, vol. 224, pp. 145--158, Aug. 2024.

\bibitem{Ntanos2020}
A.~Ntanos \emph{et~al.}, ``{QKD in Support of Secured P2P and P2MP Key Exchange for Low-Latency 5G Connectivity},'' in \emph{IEEE 3rd 5G World Forum (5GWF)}, Sep. 2020, pp. 157--162.

\bibitem{Rana2022}
V.~Rana, R.~A. Chou, and H.~M. Kwon, ``Information-theoretic secret sharing from correlated gaussian random variables and public communication,'' \emph{IEEE Transactions on Information Theory}, vol.~68, no.~1, pp. 549--559, 2022.

\bibitem{arute2019}
F.~Arute, K.~Arya, R.~Babbush, D.~Bacon, J.~C. Bardin, R.~Barends, R.~Biswas, S.~Boixo, F.~G. Brandao, D.~A. Buell \emph{et~al.}, ``Quantum supremacy using a programmable superconducting processor,'' \emph{Nature}, vol. 574, no. 7779, pp. 505--510, 2019.

\bibitem{debnath2016}
S.~Debnath, N.~M. Linke, C.~Figgatt, K.~A. Landsman, K.~Wright, and C.~Monroe, ``Demonstration of a small programmable quantum computer with atomic qubits,'' \emph{Nature}, vol. 536, no. 7614, pp. 63--66, 2016.

\bibitem{gong2021}
M.~Gong, S.~Wang, C.~Zha, M.-C. Chen, H.-L. Huang, Y.~Wu, Q.~Zhu, Y.~Zhao, S.~Li, S.~Guo \emph{et~al.}, ``Quantum walks on a programmable two-dimensional 62-qubit superconducting processor,'' \emph{Science}, vol. 372, no. 6545, pp. 948--952, 2021.

\bibitem{Cao2022}
Y.~Cao, Y.~Zhao, Q.~Wang, J.~Zhang, S.~X. Ng, and L.~Hanzo, ``The evolution of quantum key distribution networks: On the road to the qinternet,'' \emph{IEEE Communications Surveys \& Tutorials}, vol.~24, no.~2, pp. 839--894, 2022.

\bibitem{Bennet1984}
C.~H. Bennett and G.~Brassard, ``{Quantum cryptography: Public key distribution and coin tossing},'' \emph{Theoretical Computer Science}, vol. 560, pp. 7--11, Dec. 2014.

\bibitem{Ekert1991}
A.~K. Ekert, ``{Quantum cryptography based on Bell's theorem},'' \emph{Phys. Rev. Lett.}, vol.~67, pp. 661--663, Aug. 1991.

\bibitem{Grosshans2002}
F.~Grosshans and P.~Grangier, ``Continuous variable quantum cryptography using coherent states,'' \emph{Phys. Rev. Lett.}, vol.~88, p. 057902, Jan 2002.

\bibitem{Stucki2005}
D.~Stucki, N.~Brunner, N.~Gisin, V.~Scarani, and H.~Zbinden, ``Fast and simple one-way quantum key distribution,'' \emph{Applied Physics Letters}, vol.~87, no.~19, p. 194108, 11 2005.

\bibitem{Lucamarini2015}
M.~Lucamarini, J.~F. Dynes, B.~Fröhlich, Z.~Yuan, and A.~J. Shields, ``Security bounds for efficient decoy-state quantum key distribution,'' \emph{IEEE Journal of Selected Topics in Quantum Electronics}, vol.~21, no.~3, pp. 197--204, 2015.

\bibitem{Alshaer2021}
N.~Alshaer, A.~Moawad, and T.~Ismail, ``Reliability and security analysis of an entanglement-based qkd protocol in a dynamic ground-to-uav fso communications system,'' \emph{IEEE Access}, vol.~9, pp. 168\,052--168\,067, 2021.

\bibitem{10214294}
D.~Zavitsanos, A.~Ntanos, T.~Stathopoulos, A.~Raptakis, F.~Setaki, G.~Lyberopoulos, C.~Kouloumentas, G.~Giannoulis, and H.~Avramopoulos, ``Feasibility analysis of qkd integration in real-world ftth access networks,'' \emph{Journal of Lightwave Technology}, vol.~42, no.~1, pp. 4--11, 2024.

\bibitem{Bahrami2020}
A.~Bahrami, A.~Lord, and T.~Spiller, ``{Quantum key distribution integration with optical dense wavelength division multiplexing: a review},'' \emph{IET Quantum Communication}, vol.~1, no.~1, pp. 9--15, Jul. 2020.

\bibitem{Biswas2021}
A.~Biswas, A.~Banerji, P.~Chandravanshi, R.~Kumar, and R.~P. Singh, ``Experimental side channel analysis of bb84 qkd source,'' \emph{IEEE Journal of Quantum Electronics}, vol.~57, no.~6, pp. 1--7, 2021.

\bibitem{Burdiak2023}
P.~Burdiak, L.~Kapičák, L.~Michalek, E.~Dervisevic, M.~Mehic, and M.~Vozňák, ``Demonstration of qkd integration into 5g campus network,'' in \emph{International Conference on Software, Telecommunications and Computer Networks (SoftCOM)}, 2023, pp. 1--4.

\bibitem{Chen2009}
W.~Chen, Z.-F. Han, T.~Zhang, H.~Wen, Z.-Q. Yin, F.-X. Xu, Q.-L. Wu, Y.~Liu, Y.~Zhang, X.-F. Mo, Y.-Z. Gui, G.~Wei, and G.-C. Guo, ``Field experiment on a “star type” metropolitan quantum key distribution network,'' \emph{IEEE Photonics Technology Letters}, vol.~21, no.~9, pp. 575--577, 2009.

\bibitem{Tessinari2023}
R.~S. Tessinari, R.~I. Woodward, and A.~J. Shields, ``Software-defined quantum network using a qkd-secured sdn controller and encrypted messages,'' in \emph{Optical Fiber Communications Conference and Exhibition (OFC)}, 2023, pp. 1--3.

\bibitem{Aji2021}
A.~Aji, K.~Jain, and P.~Krishnan, ``{A Survey of Quantum Key Distribution (QKD) Network Simulation Platforms},'' in \emph{2nd Global Conference for Advancement in Technology (GCAT)}, Oct. 2021, pp. 1--8.

\bibitem{Dervisevic2024}
E.~Dervisevic, M.~Voznak, and M.~Mehic, ``{Large-scale quantum key distribution network simulator},'' \emph{Journal of Optical Communications and Networking}, vol.~16, no.~4, pp. 449--462, Apr. 2024.

\bibitem{Ahmed2012}
A.~I. Khaleel, ``{Coherent one-way protocol: Design and simulation},'' in \emph{International Conference on Future Communication Networks}, Apr. 2012, pp. 170--174.

\bibitem{Stucki_2009}
D.~Stucki \emph{et~al.}, ``{Continuous high speed coherent one-way quantum key distribution},'' \emph{Opt. Express}, vol.~17, no.~16, pp. 13\,326--13\,334, Aug. 2009.

\bibitem{Buttler1998}
W.~T. Buttler \emph{et~al.}, ``{Practical Free-Space Quantum Key Distribution over 1 km},'' \emph{Phys. Rev. Lett.}, vol.~81, pp. 3283--3286, Oct. 1998.

\bibitem{kato1972schrodinger}
T.~Kato, ``{Schr{\"o}dinger operators with singular potentials},'' \emph{Israel Journal of Mathematics}, vol.~13, pp. 135--148, Mar. 1972.

\bibitem{Gao2022}
R.-Q. Gao \emph{et~al.}, ``{Simple security proof of coherent-one-way quantum key distribution},'' \emph{Opt. Express}, vol.~30, no.~13, pp. 23\,783--23\,795, Jun. 2022.

\bibitem{Lodewyck2007}
J.~Lodewyck, M.~Bloch, R.~Garc\'{\i}a-Patr\'on, S.~Fossier, E.~Karpov, E.~Diamanti, T.~Debuisschert, N.~J. Cerf, R.~Tualle-Brouri, S.~W. McLaughlin, and P.~Grangier, ``Quantum key distribution over $25\phantom{\rule{0.3em}{0ex}}\mathrm{km}$ with an all-fiber continuous-variable system,'' \emph{Phys. Rev. A}, vol.~76, p. 042305, Oct 2007.

\bibitem{Tomamichel2011}
M.~Tomamichel and R.~Renner, ``{Uncertainty Relation for Smooth Entropies},'' \emph{Phys. Rev. Lett.}, vol. 106, p. 110506, Mar. 2011.

\bibitem{Konig2009}
R.~Konig, R.~Renner, and C.~Schaffner, ``{The Operational Meaning of Min- and Max-Entropy},'' \emph{IEEE Transactions on Information Theory}, vol.~55, no.~9, pp. 4337--4347, Sep. 2009.

\bibitem{Renes2012}
J.~M. Renes and R.~Renner, ``{One-Shot Classical Data Compression With Quantum Side Information and the Distillation of Common Randomness or Secret Keys},'' \emph{IEEE Transactions on Information Theory}, vol.~58, no.~3, pp. 1985--1991, Mar. 2012.

\bibitem{Tomamichel2012}
M.~Tomamichel, C.~C.~W. Lim, N.~Gisin, and R.~Renner, ``{Tight finite-key analysis for quantum cryptography},'' \emph{Nature Communications}, vol.~3, no. 634, Jan. 2012.

\bibitem{Moroder2012}
T.~Moroder \emph{et~al.}, ``{Security of Distributed-Phase-Reference Quantum Key Distribution},'' \emph{Phys. Rev. Lett.}, vol. 109, p. 260501, Dec. 2012.

\bibitem{Wang2019}
Y.~Wang \emph{et~al.}, ``{Characterising the correlations of prepare-and-measure quantum networks},'' \emph{npj Quantum Information}, vol.~5, no.~17, Feb. 2019.

\bibitem{Li2011}
H.-W. Li \emph{et~al.}, ``{Attacking a practical quantum-key-distribution system with wavelength-dependent beam-splitter and multiwavelength sources},'' \emph{Phys. Rev. A}, vol.~84, p. 062308, Dec. 2011.

\bibitem{Li2015}
------, ``{Randomness determines practical security of BB84 quantum key distribution},'' \emph{Scientific Reports}, vol.~5, no. 16200, Nov. 2015.

\bibitem{Sun2020}
S.-H. Sun \emph{et~al.}, ``{Security evaluation of quantum key distribution with weak basis-choice flaws},'' \emph{Scientific Reports}, vol.~10, no. 18145, Oct. 2020.

\bibitem{Islam2017}
N.~T. Islam \emph{et~al.}, ``{Provably secure and high-rate quantum key distribution with time-bin qudits},'' \emph{Science Advances}, vol.~3, no.~11, p. e1701491, Nov. 2017.

\bibitem{Chen2021}
Y.~Chen \emph{et~al.}, ``{An integrated space-to-ground quantum communication network over 4,600 kilometres},'' \emph{Nature}, vol. 589, pp. 214--219, Jan. 2021.

\bibitem{Beaudry2008}
N.~J. Beaudry, T.~Moroder, and N.~L\"utkenhaus, ``{Squashing Models for Optical Measurements in Quantum Communication},'' \emph{Phys. Rev. Lett.}, vol. 101, p. 093601, Aug. 2008.

\bibitem{Cao2016}
Z.~Cao, H.~Zhou, X.~Yuan, and X.~Ma, ``{Source-Independent Quantum Random Number Generation},'' \emph{Phys. Rev. X}, vol.~6, p. 011020, Feb. 2016.

\bibitem{Koashi2009}
M.~Koashi, ``{Simple security proof of quantum key distribution based on complementarity},'' \emph{New Journal of Physics}, vol.~11, no.~4, p. 045018, Apr. 2009.

\bibitem{curty2018}
M.~Curty, K.~Azuma, and H.-K. Lo, ``{Simple security proof of twin-field type quantum key distribution protocol},'' \emph{npj Quantum Information}, vol.~5, no.~64, Jul. 2019.

\bibitem{Rust}
``{Rust Programming Language},'' \url{https://www.rust-lang.org/}, accessed: 2025-01-14.

\bibitem{Actix}
``{Actix Web},'' \url{https://actix.rs/}, accessed: 2025-01-14.

\bibitem{PostgreSQL}
``{PostgreSQL: The World's Most Advanced Open Source Relational Database},'' \url{https://www.postgresql.org/}, accessed: 2025-01-14.

\bibitem{WebAssembly}
``{WebAssembly},'' \url{https://webassembly.org/}, accessed: 2025-01-14.

\bibitem{Trunk}
``{Trunk},'' \url{https://trunkrs.dev/}, accessed: 2025-01-14.

\bibitem{YEW-RS}
``{YEW-RS},'' \url{https://yew.rs/}, accessed: 2025-01-14.

\bibitem{TailwindCSS}
``{Tailwind CSS - Rapidly build modern websites without ever leaving your HTML.}'' \url{https://tailwindcss.com/}, accessed: 2025-01-14.

\bibitem{ETSI014}
\BIBentryALTinterwordspacing
ETSI, ``{GS QKD 014 Quantum Key Distribution: Protocol and data format of REST-based key delivery API},'' Feb. 2019. [Online]. Available: \url{https://www.etsi.org/deliver/etsi_gs/QKD/001_099/014/01.01.01_60/gs_qkd014v010101p.pdf}
\BIBentrySTDinterwordspacing

\bibitem{IDQuantique}
``{ID Quantique - The home of Quantum-Safe Crypto},'' \url{https://www.idquantique.com/}, accessed: 2025-01-14.

\bibitem{TelloCastillo}
A.~T. Castillo, E.~Eso, and R.~Donaldson, ``In-lab demonstration of coherent one-way protocol over free space with turbulence simulation,'' \emph{Opt. Express}, vol.~30, no.~7, pp. 11\,671--11\,683, Mar 2022.

\bibitem{Stucki_2009b}
D.~Stucki, N.~Walenta, F.~Vannel, R.~T. Thew, N.~Gisin, H.~Zbinden, S.~Gray, C.~R. Towery, and S.~Ten, ``High rate, long-distance quantum key distribution over 250 km of ultra low loss fibres,'' \emph{New Journal of Physics}, vol.~11, no.~7, p. 075003, jul 2009.

\end{thebibliography}

\vfill

\end{document}